\documentclass[aps,superscriptaddress, showpacs,preprintnumbers, superscriptaddress, nofootinbibt, twocolumn]{revtex4-2}
\usepackage{eurosym}
\usepackage{amsfonts}
\usepackage{amssymb,amsmath,enumitem}
\usepackage{graphicx,color,epstopdf}
\usepackage{subfigure}
\usepackage{ulem}
\usepackage{tabularx}
\usepackage{booktabs}

\def\cL{\mathcal{L}}

\def\tn{\tilde{\nabla}}
\def\tg{\tilde{\Gamma}}
\def\be{\begin{equation}}
\def\ee{\end{equation}}
\def\beg{\begin{align}}
\def\eeg{\end{align}}
\def\bea{\begin{eqnarray}}
\def\eea{\end{eqnarray}}

\def\nn{\nonumber \\}

\def\l{\left}
\def\r{\right}
\newcommand{\f}[2]{\frac{#1}{#2}}

\begin{document}

\title{Geodesic deviation, Raychaudhuri equation,  Newtonian limit, and tidal forces in Weyl-type $f(Q,T)$ gravity}
\author{Jin-Zhao Yang}
\email{yangjch6@mail3.sysu.edu.cn}
\affiliation{School of Physics, Sun Yat-Sen University, Xingang Road, Guangzhou 510275,
P. R. China,}
\author{Shahab Shahidi}
\email{s.shahidi@du.ac.ir}
\affiliation{School of Physics, Damghan University, Damghan,
	41167-36716, Iran}
\author{Tiberiu Harko}
\email{tiberiu.harko@aira.astro.ro}
\affiliation{Astronomical Observatory, 19 Ciresilor Street,  Cluj-Napoca 400487, Romania,}
\affiliation{Department of Physics, Babes-Bolyai University, Kogalniceanu Street,
Cluj-Napoca 400084, Romania}
\affiliation{School of Physics, Sun Yat-Sen University, Xingang Road, Guangzhou 510275,
P. R. China,}
\author{Shi-Dong Liang}
\email{stslsd@mail.sysu.edu.cn}
\affiliation{School of Physics, Sun Yat-Sen University, Xingang Road, Guangzhou 510275,
P. R. China,}
\affiliation{State Key Laboratory of Optoelectronic Material and Technology,
and Guangdong Province Key Laboratory of Display Material and Technology,
Sun Yat-Sen University, Guangzhou 510275, People’s Republic of China.}
\date{\today}

\begin{abstract}
We consider the geodesic deviation equation, describing the relative accelerations of nearby particles, and the Raychaudhuri equation, giving the evolution of the kinematical quantities associated with deformations
(expansion, shear and rotation) in the Weyl-type $f(Q, T)$ gravity, in which the non-metricity $Q$ is
represented in the standard Weyl form, fully determined by the Weyl vector, while $T$ represents the trace of
the matter energy–momentum tensor. The effects of the Weyl geometry and of the extra force induced by
the non-metricity–matter coupling are explicitly taken into account. The Newtonian limit of the theory is
investigated, and the generalized Poisson equation, containing correction terms coming from the Weyl
geometry, and from the geometry matter coupling, is derived. As a physical application of the geodesic
deviation equation the modifications of the tidal forces, due to the non-metricity–matter coupling, are
obtained in the weak-field approximation. The tidal motion of test particles is directly influenced by the
gradients of the extra force, and of the Weyl vector. As a concrete astrophysical example we obtain the
expression of the Roche limit (the orbital distance at which a satellite begins to be tidally torn apart by the
body it orbits) in the Weyl-type $f(Q, T)$ gravity.
\end{abstract}
\pacs{03.75.Kk, 11.27.+d, 98.80.Cq, 04.20.-q, 04.25.D-, 95.35.+d}
\maketitle
\tableofcontents

\section{Introduction}

The twentieth century has seen the birth of General Relativity (GR) and of Quantum Mechanics (QM), which are considered as the most two successful theories describing the nature and properties of the physical world, on scale ranging from the microscopic to the cosmological one.  General Relativity, a geometrical theory of gravity, one of the fundamental interactions that shape the Universe, has provided an excellent description of the observational data \cite{will}, and has led to new insights into the problems of space and time, and their relation with the physical Universe. The recent detection \cite{abbott} of the gravitational waves has again proved the existence of the excellent correspondence between experimental data and the theoretical predictions of GR in a range extending from weak to strong gravitational fields. The introduction of Riemann geometry into GR has provided a powerful mathematical framework to describe the properties of the gravitational field.  However, despite its remarkable success,  a number of recent observational results have raised some questions about the absolute validity of standard GR, which may still present some limitations especially on astrophysical scales exceeding the Solar System one. The two fundamental problems facing present day gravitational theories, and, in particular General Relativity, are the dark energy and the dark matter problems, respectively. For recent reviews on the dark energy and dark matter problems see \cite{RDE0, RDE1,RDE2,RDE3,RDE4,RDE5,RDM1,RDM2,RDM3,RDM4,RDM5}. To fix the theoretical imperfections of standard General Relativity, two important physical components, the dark matter and the dark energy, are introduced in the cosmological scenario in a rather ad hoc way, with the major goal of explaining and adapting the standard gravity model to describe realistic physical situations, related to the motion  of massive particles around galaxies, and to solve the problem of the accelerated expansionary state of Universe. We may call the approach based on the introduction of two new physical components in the overall matter/energy balance of the Universe as the dark components model \cite{Bey}.

However, a second approach to gravitational phenomena is also possible, and it is called the dark gravity approach \cite{Bey}. In this approach one assumes that both dark matter and dark energy can be explained by changing the nature of the gravitational force.  Many modified theory of gravity, going beyond the standard GR model have been proposed in order to build a fundamental framework for the explanation mysterious dark matter and dark energy \cite{avelino},  and to give a solution of the present observational-theoretical contradictions and conflicts. In the framework of the Riemannian geometry one can naturally generalize the Einstein-Hilbert action by substituting the Ricci scalar $R$ with an arbitrary functions $f(R)$. This leads to the $f(R)$ modified theory of gravity \cite{nojiri2011}-\cite{capo2008}. There are two approaches in $f(R)$ gravity, namely, the metric formulation, in which the metric is considered as the only dynamical variable, and the Palatini formulation, in which the connection is considered as another independent variable, beside the metric tensor. One can find detailed discussions of the metric formulation of $f(R)$ gravity in \cite{sotiriou2006}-\cite{lobo1}, and for the Palatini formulation in \cite{li2007}-\cite{stachowski2017}, respectively. The most obvious drawbacks of $f(R)$ theory is that the scalar field in the Palatini formulation is not dynamical, which implies that no new degrees of freedom can be introduced,  resulting into the existence of infinite tidal forces that generally are physically impossible \cite{olmo2011}. On the other hand the extra freedom introduced from the metric formulation would lead to contradiction with the observational results obtained in the Solar System \cite{capo20083,khoury}.

To allow for the generation of long-range forces and simultaneously passing the Solar system test, in \cite{harko20122}-\cite{capo2209} a new approach to gravitational effects was proposed. In this theory, called Hybrid Metric-Palatini Gravity,  the Einstein-Hilbert action is supplemented with a correction term inspired by the Palatini formulation. Another interesting and important modification of gravity is the inclusion of a non-minimal coupling of geometry and matter into the  action \cite{berto2007}-\cite{fRT}, by using arbitrary functions of the scalar curvature and Lagrangian density of matter (in the $f\left(R,L_m\right)$ gravity theory \cite{fRLm}), or by considering a gravitational Lagrangian of the form $f(R,T)$ \cite{fRT}, where $T$ is the trace of the matter energy-momentum tensor. In these classes of theories the covariant derivative of the energy-momentum tensor is always non-zero, which implies a non-geodesic motion of test particles, and the appearance of  an extra force. For a  recent review of  some modified gravity theories in Riemann geometry one can refer to \cite{Bey} and \cite{harko2018}, respectively.

The standard GR theory is formulated in Riemann geometry. Hence, an alternative avenue for searching for a generalized description of gravity is the extensions of the geometrical framework on which GR is based. In an attempt to unify gravity and electromagnetism  H. Weyl introduced in 1918 a generalization of the Riemann geometry \cite{weyl1918}. In the Weyl geometry both the orientation and the length of vectors are allowed to vary under parallel transport, while in Riemann geometry only the variation of the orientation is allowed. Weyl geometry represents a completely consistent generalization of Riemannian geometry. In modern language, the vector field introduced by Weyl, which generates a new component in the connection, which in Weyl geometry is no longer metric-compatible, is actually the dilatation gauge vector. If the vector is the gradient of some function, a scale transformation of the form $\tilde g_{\mu\nu} = \sigma^2 g_{\mu\nu}$, where $\sigma$ is the scale (conformal) factor, can be applied to cancel the Weyl vector. In this case the Weyl geometry is called integrable, and the length of vectors will be unchanged under a parallel transport along closed paths. For a detailed discussion of Weyl geometry see \cite{adler,adler1,adler2}.

In an important mathematical and physical development E. Cartan introduced the anti-symmetric part of the connection, known as torsion, into the gravity theory,  thus formulating  an extension of GR \cite{cartan1979}, which is known as the Einstein-Cartan theory. The Weyl geometry can be immediately generalized by including the torsion, which leads to  the Weyl-Cartan geometry, and a corresponding geometric theory of gravitation \cite{borze1,borze2,borze3,borze4,borze5,borze6,borze7,borze8,borze9,borze10,borze11, borze12}. For a review of geometrical properties and physical applications of Riemann-Cartan and Weyl-Cartan spacetimes one can refer to \cite{nove2008}. Another important mathematical development with important physical implications is related to the work by R. Weitzenb{\"o}ck \cite{weizen1923}, who developed a geometry with torsion and zero Riemann curvature. Since curvature is zero, Weizenb\"ock spaces possess the interesting property of distant parallelism, known as teleparallelism. The teleparallel approach  substitutes the metric tensor, which plays  a central role as a basic physical variable in gravitational theories, with a set of tetrad vectors. In this approach, the torsion is generated by tetrad fields and it describes the gravitational field entirely, once the torsion is properly chosen to eliminate the curvature. This is the so-called teleparallel equivalent of general relativity (TEGR) \cite{moller1, moller2, moller3}, which is also known as $f(\tilde{T})$ theory, where $\tilde{T}$ is the trace of torsion tensor. In $f(\tilde{T})$ theory, torsion completely compensates the curvature, and the spacetime becomes flat. Weyl-Cartan-Weizenb\"ock gravity (WCW) was introduced as an extension of the teleparallel gravity models in \cite{hag2012}. In this approach, the Weizenb\"ock condition of the vanishing of the total curvature is implemented in a Weyl-Cartan spacetime. Moreover, in \cite{hag2013} the Weizenb\"ock condition of the exact compensation of torsion and curvature was introduced in the action via the Lagrange multiplier approach, in Riemann-Cartan spacetime. For a review of teleparallel gravity theory see \cite{cai2016}.

An interesting theory, geometrically  equivalent to GR,  which is also known as symmetric teleparallel gravity, was introduced in 1999 \cite{nester1999}. In this geometric approach the nonmetricity $Q$ of a Weyl geometry represents the basic geometrical variable. This approach was further developed as $f(Q)$ gravity theory, also named as nonmetric gravity \cite{beltran2018}. Various physical and geometrical properties of symmetric teleparallel gravity have been analyzed in \cite{s1,s2,s3,s4,s5,s6,s7,s8,s9, s11, s12a, s13,s14,s15,s16,s17,s18,s19,s20,s21}.

An important extension of $f(Q)$ theory has been obtained in \cite{fQ1} by including a nonminimal curvature-matter coupling into the gravitational Lagrangian, with $L = f_1(Q) + f_2(Q) \cL_m$, where $f_1$ and $f_2$ are arbitrary functions of the nonmetricity $Q$. Similarly to the nonminimal couplings between curvature and matter in Riemannian geometry \cite{fRLm,fRT}, the coupling leads to the non-conservation of the matter energy-momentum tensor as well, which leads to the existence of a new term in the geodesic equation, which can be interpreted as an extra-force. A Bayesian statistical analysis using redshift space distortions data was performed to test a model of $f(Q)$ gravity in \cite{fQ2}. The cosmological background evolution is similar to the  $\Lambda$CDM one, but differences arise in the perturbations. The best fit parameters indicate that the $\sigma8$ tension between Planck and Large Scale Structure data can be relieved in this theory.

The $f(Q)$ gravity model was extended to include a nonminimal coupling in the Lagrangian in \cite{fQT}, in which the gravitational action $L$ is given by an arbitrary function $f$ of the nonmetricity $Q$ and of the trace of the matter energy-momentum tensor $T$, with $L=f(Q,T)$. Several cosmological applications of the theory were considered by chosing some simple functional forms of the function $f(Q,T)$, corresponding to additive expressions of $f(Q,T)$ of the form $f(Q,T)=\alpha Q+\beta T$, $f(Q,T)=\alpha Q^n+1+\beta T$, and $f(Q,T)=-\alpha Q-\beta T^2$, respectively, where $\alpha$, $\beta $ and $n$ are constants. The Hubble function, the deceleration parameter, and the matter energy density were obtained in each case as a function of the redshift by using analytical and numerical techniques. For all considered cases the Universe experiences an accelerating expansion, ending with a de Sitter type evolution. Gravitational baryogenesis in $f(Q,T)$ gravity was considered in \cite{fQT1}, and it was found that $f(Q,T)$ gravity can contribute significantly to this phenomenon.  The various cosmological parameters in Friedmann-Lemaitre-Robertson-walker (FLRW) geometry have been obtained in \cite{fQT2} for different choices of the function $f(Q,T)$ in terms of the scale-factor and redshift $z$ by constraining the energy-conservation law. The observational constraints on the model have been obtained by fitting the model parameters using the available data sets like Hubble data sets $H(z)$, Joint Light Curve Analysis (JLA) data sets and union 2.1 compilation of SNe Ia data sets. The various energy conditions for cosmological models in $f(Q,T)$ gravity were studied in \cite{fQT3}.   The equation of state parameter $w=-1$  also supports the accelerating behavior of the Universe. In the considered $f(Q,T)$ models  the null, weak, and dominant energy conditions are obeyed, while the strong energy conditions are violated during the present accelerated expansion.  The late time cosmology in $f(Q,T)$ gravity was investigated in \cite{fQT4}. Constraints on the model parameters were imposed from the updated 57 points of Hubble data sets, and 580 points of union 2.1 compilation supernovae data sets. The performed analysis did show that $f(Q,T)$ gravity represents a promising approach for explaining the current cosmic acceleration,  and it can provide a consistent solution to the dark energy problem.

A particular type of $f(Q,T)$ model was considered in \cite{Xu2020}, in which the scalar non-metricity $Q_{\mu \nu}$ of the space-time was expressed in its standard Weyl form, and therefore it is fully determined by a vector field $w_{\mu}$. The field equations of the theory have been obtained under the assumption of the vanishing of the total scalar curvature, a condition which was added into the gravitational action via a Lagrange multiplier. The cosmological implications of the theory were also investigated  for a flat, homogeneous and isotropic geometry, and the generalized Friedmann equations were obtained. Several cosmological models were investigated by adopting some simple functional forms of the function $f(Q, T)$, and  the predictions of the theory have been compared with the standard $\Lambda$CDM model.

The main goal of the present paper is to investigate some fundamental properties of motion of test particles in the Weyl type $f(Q,T)$ gravity theory, introduced in  \cite{Xu2020}. We derive the geodesic deviation equation and the Raychaudhuri equation in the Weyl geometry and in the presence of the extra-force that describes the effects of the nonmetricity-matter coupling. As compared to standard general relativity a number of new terms do appear in both equations, indicating the existence of a complex dynamics resulting from the intricate interplay of the geometrical and matter factors. The weak field limit of the theory is also considered, and the Poisson equation, containing the corrections arriving from the Weyl geometry and nonmetricity-matter coupling are determined. As a physical application of the obtained results we consider the tidal force problem in Weyl type $f(Q,T)$ gravity, and a generalization of the Roche limit is obtained,  under the simplifying assumption that the center of mass of the two-body system coincides with the geometrical center of the massive object with mass $M$.

The present paper is organized as follows. In Section~\ref{sect1} we review the basics of the Weyl geometry, and we introduce the Weyl type $f(Q,T)$ gravity theory, and its field equations. The Geodesic deviation equation is derived in Section~\ref{sect2}, while the Raychaudhuri equation in Weyl geometry and in the presence of the extra force induced by the nonmetricity-matter coupling is obtained in Section~\ref{sect3}. In Section~\ref{sect4} we consider the weak field limit of the theory, and the generalized Poisson equation for the gravitational potential, containing the corrections due to the geometric and coupling effects, is obtained.  The properties of the tidal forces, as well as the expression of the Roche limit in Weyl geometry and in the presence of extra force, are also investigated. We  discuss and conclude our results in Section~\ref{sect5}. Some mathematical results used in the derivation of the main relations of the paper are summarized and detailed in the Appendix~\ref{app}.

\section{Geometrical preliminaries and the basics of the Weyl-type $f(Q,T)$ gravity theory}\label{sect1}

In the present Section we briefly review the fundamentals of the Weyl geometry, and of the Weyl-type $f(Q,T)$ gravity.

\subsection{Quick start for Weyl geometry}

In the differential geometry of the Riemann spaces the connection describes the properties of the parallel transport, the fundamental concept used for the characterization of the various mathematical aspects associated to the geometrical objects in curved space-times. In order to keep the tensor properties of the differentiation operation acting on a vector, some new terms, containing the connection, must be introduced to assure the tensorial nature of the differentials of a vector. Hence, in Riemann geometry  the covariant derivatives of the contravariant and covariant vectors are given by \cite{landau,gron,hans}
\begin{eqnarray}\label{weyl1}
\hspace{-0.3cm}\nabla_\nu A^\mu = \partial_\nu A^\mu + \Gamma^\mu_{\ \lambda\nu} A^\lambda, \nabla_\nu A_\mu = \partial_\nu A_\mu - \Gamma^\lambda_{\ \mu\nu} A_\lambda,
\end{eqnarray}

In Riemann spaces, the connection is of Christoffel type, it is compatible to the metric $g_{\mu \nu}$, so that $\nabla _{\lambda}g_{\mu \nu}=0$, and it is given by
\begin{eqnarray}\label{weyl2}
\Gamma^\lambda_{\ \mu\nu} = \frac{1}{2} g^{\lambda\sigma} \bigg( \partial_\mu g_{\nu\sigma} + \partial_\nu g_{\mu\sigma} - \partial_\sigma g_{\mu\nu} \bigg).
\end{eqnarray}

With this connection the parallel transportation in a Riemann space will preserve the length of the vectors,  and only the orientation of the vector is changed. In 1918 H. Weyl proposed a new geometry \cite{weyl1918} by introducing a connection with the property that under parallel transportation both the orientation and the magnitude of a vector change. The connection in Weyl geometry is no longer metric compatible. Moreover, a new vector field, known as Weyl vector field, is introduced,  allowing to write the Weyl connection as \cite{adler1,adler2},
\begin{eqnarray}\label{weyl3}
\tilde{\Gamma}^\lambda_{\ \mu\nu} = \Gamma^\lambda_{\ \mu\nu} + g_{\mu\nu}w^\lambda-\delta^\lambda_{\mu} w_\nu - \delta^\lambda_\nu w_\mu,
\end{eqnarray}
where in the following the tilde indicates the quantities defined in the Weyl geometry. With the use of the Weyl connection one can immediately obtain the fundamental result that the covariant derivative associated with the connection $\tilde{\Gamma}^\lambda_{\ \mu\nu}$, when applied on the metric tensor, gives a non-zero value,
\begin{eqnarray}\label{weyl4}
\tn_\lambda g_{\mu\nu} = 2w_\lambda g_{\mu\nu},\quad \tn_\lambda g^{\mu\nu} = -2w_\lambda g^{\mu\nu}.
\end{eqnarray}

In the standard representation of the Weyl geometry the variation of the length of a vector under parallel transport is given by
\begin{eqnarray}\label{weyl5}
\delta l = l_0 w_\mu \delta x^\mu,
\end{eqnarray}
where $l_0$ is the length of the vector before transportation. Thus the variation of the length of a vector under transport along a closed loop is,
\begin{eqnarray}\label{weyl6}
\delta l = l_0 W_{\mu\nu} \delta s^{\mu\nu},
\end{eqnarray}
where $\delta s^{\mu\nu}$ is the area surrounded by the loop, and $W_{\mu\nu}$ is given by
\begin{eqnarray}\label{weyl7}
W_{\mu\nu} = \tn_\nu w_\mu - \tn_\mu w_\nu = \nabla_\nu w_\mu - \nabla_\mu w_\nu.
\end{eqnarray}

In Weyl geometry one can define a curvature tensor in the same way as in Riemann geometry, with the use of the intrinsic connection,
\begin{eqnarray}\label{weyl8}
2\tilde \nabla_{[\mu} \tilde \nabla_{\nu]} A_\lambda &=& \tilde R^\sigma_{\ \lambda\nu\mu} A_\sigma,\\
2\tilde \nabla_{[\mu} \tilde \nabla_{\nu]} A^\lambda &=& -\tilde R^\lambda_{\ \sigma\nu\mu} A^\sigma.
\end{eqnarray}

The explicit expressions of the curvature tensor  in Weyl geometry can be obtained as
\begin{eqnarray}\label{weyl9}
\hspace{-0.1cm}&& \tilde R_{\mu\nu\lambda\sigma}
= R_{\mu\nu\lambda\sigma} + g_{\mu\nu}W_{\lambda\sigma} + 2\nabla_\lambda w_{[\mu}g_{\nu]\sigma} + 2\nabla_\sigma w_{[\nu}g_{\mu]\lambda} +\nn
\hspace{-0.1cm}&& 2w_\lambda w_{[\mu}g_{\nu]\sigma} + 2 w_\sigma w_{[\nu}g_{\mu]\lambda} - 2w^\alpha w_\alpha g_{\lambda[\mu}g_{\nu]\sigma}.
\end{eqnarray}
By contracting the first and the third indices of the  curvature tensor, we find
\begin{eqnarray}\label{weyl10}
\tilde R_{\nu\sigma} &=& R_{\nu\sigma} + 2w_\nu w_\sigma + 2\nabla_\sigma w_\nu +W_{\nu\sigma}+\nn
&& g_{\nu\sigma} (\nabla_\alpha w^\alpha - 2w_\alpha w^\alpha).
\end{eqnarray}

Finally, contracting the remaining two indices one can obtain the scalar curvature as
\begin{eqnarray}\label{weyl11}
\tilde R = \tilde R_\mu^{\ \mu} = R + 6(\nabla_\mu w^\mu - w_\mu w^\mu).
\end{eqnarray}

In order to develop some physical applications we introduce two types of nonmetricities as follows,
\begin{eqnarray}\label{weyl12}
Q_{\lambda\mu\nu} &=& - 2w_\lambda g_{\mu\nu} = - \tilde \nabla_\lambda g_{\mu\nu},\\
Q^{\lambda\mu\nu} &=& - 2w^\lambda g^{\mu\nu} = \tn^\lambda g^{\mu\nu}.
\end{eqnarray}

Moreover, the scalar nonmetricity $Q$ is defined as
\begin{equation}\label{weyl13}
Q = -g^{\mu\nu} \l(L^\alpha_{~\beta\mu}L^\beta_{\nu\alpha}-L^{\alpha}_{~\beta\alpha}L^\beta_{\mu\nu}\r),
\end{equation}
where $L^\lambda_{\ \mu\nu}$ is given by
\begin{equation}\label{weyl14}
L^\lambda_{\ \mu\nu} = \frac{1}{2}g^{\lambda\sigma}\l(Q_{\mu\sigma\nu} + Q_{\nu\sigma\mu} - Q_{\sigma\mu\nu}\r).
\end{equation}

 Hence we can rewrite the connection in Weyl geometry as
 \be
 \tilde \Gamma^\lambda_{\ \mu\nu} = \Gamma^\lambda_{\ \mu\nu} + L^\lambda_{\ \mu\nu}.
 \ee

 With the use of Eqs.~(\ref{weyl12}) and (\ref{weyl14}), we obtain for the scalar nonmetricity the expression (see Appendix~\ref{app1} for the calculational details),
\begin{equation}\label{weyl15}
Q = 6w^2.
\end{equation}

The explicit expressions of the covariant derivatives for some vector and tensor quantities in Weyl geometry are given in Appendix~\ref{app2}.

\subsection{The Weyl-type  $f(Q,T)$ gravity theory}

A particular model of the general $f(Q,T)$ theory was introduced in \cite{Xu2020}, and it is based on the following action,
\begin{eqnarray}\label{fqt1}
S &=& \int d^4x\sqrt{-g}\bigg[\kappa^2 f(Q,T) - \frac{1}{4}W_{\mu\nu}W^{\mu\nu} -\nonumber\\
&& \frac{1}{2}m^2 w_\mu w^\mu + \lambda\tilde R + \cL_m \bigg],
\end{eqnarray}
where we have denoted $\kappa ^2=1/16\pi G$, and $m$ is the mass of the particle
associated to the vector field $w_{\mu}$. Moreover,  with $\mathcal{L}_m$ we have denoted the ordinary matter
action. In the action (\ref{fqt1}), the second and third terms are the  kinetic energy and the mass terms associated to the Weyl-type vector field,
respectively. As for the first term, giving the gravitational Lagrangian, it is taken as an arbitrary function of the nonmetricity $Q$ and of the trace $T$ of the matter energy-momentum tensor.

Moreover, in the present approach,  we assume the vanishing of the total scalar curvature, and we impose in the total action the Weizenb\"ock condition with the help of the Lagrange multiplier $\lambda$. We impose this condition in order to follow the essential ideas of the teleparallel approach to gravity. The  action (\ref{fqt1}) generalizes the GR equivalent symmetric teleparallel gravity theory, or $f(Q)$ theory, by introducing a matter-geometry coupling, in a similar manner as in the $f(R,T)$ theory, and adding an extra matter distribution for the particles associated with the Weyl vector field. However, it should be mentioned that the present theory is different from $f(R,T)$ theory due to the presence of a boundary term, and hence it can be seen as a generalized equivalent theory of GR, in the sense discussed in \cite{s13}.

	 In the case where $f(Q,T)=Q$, and if we have a vanishing mass $m=0$, the theory reduces to the symmetric telleparallel equivalent to GR, as is fully developed in \cite{beltran2018}.

By first varying the total action (\ref{fqt1}) with respect to the vector field $w^\mu$, we obtain the generalized Proca equation that describes the evolution of the vector field $w_{\mu}$ in the Weyl geometry,
\begin{equation}\label{fqt2}
\nabla^\nu W_{\mu\nu} - \l(m^2+12\kappa^2f_Q + 12\lambda\r)w_{\mu} = 6\nabla_\mu\lambda,
\end{equation}
where $f_Q=\partial f(Q,T)/\partial Q$. Next, we vary the action with respect to the metric tensor, and with the use of the Weizenb\"ock condition, that is, after omitting the terms that explicitly contain $\tilde R$, we obtain the generalized field equation of the Weyl-type $f(Q,T)$ gravity as follows,
\begin{align}\label{fqt3}
&\frac{1}{2}\l(T_{\mu\nu}+S_{\mu\nu}\r)-\kappa^2f_T\l(T_{\mu\nu}+\Theta_{\mu\nu}\r) \nn
&= -\frac{1}{2}\kappa^2fg_{\mu\nu} + 6\kappa^2f_Qw_\mu w_\nu +\nn
& \quad \lambda\l(R_{\mu\nu}-6w_\mu w_\nu + 3g_{\mu\nu} \nabla_\rho w^\rho\r) +\nn
& \quad 3g_{\mu\nu} w^\rho\nabla_\rho \lambda- 6w_{(\mu}\nabla_{\nu)}\lambda+g_{\mu\nu}\Box\lambda - \nabla_\mu\nabla_\nu \lambda,
\end{align}
where in the field equations we have denoted by $f_T$ the partial derivative of $f$ with respect to $T$, $f_T = \partial f/\partial T$, and where we have introduced the energy-momentum tensor $S_{\mu \nu}$ associated to the Weyl type vector field, defined according to
\begin{eqnarray}\label{fqt4}
S_{\mu\nu} &=& -\frac{1}{4}g_{\mu\nu}W_{\alpha\beta}W^{\alpha\beta} + W_{\mu\rho}W_\nu^{\ \rho} -\nn
&& \frac{1}{2}m^2g_{\mu\nu}w_\rho w^\rho + m^2w_\mu w_\nu.
\end{eqnarray}
 As customary, the  energy-momentum tensor $T_{\mu\nu}$ of the ordinary matter is defined as
\begin{equation}\label{fqt5}
T_{\mu\nu} = -\frac{2}{\sqrt{-g}}\frac{\delta(\sqrt{-g}\cL_m)}{\delta g^{\mu\nu}}.
\end{equation}
The tensor $\Theta_{\mu\nu}$ is obtained from the variation of the energy-momentum tensor with respect to the metric. By adopting the assumption according to which the ordinary matter Lagrangian density $\cL_m$ depends only on the metric tensor components, and not on their derivatives, we obtain for $\Theta_{\mu\nu}$ the simple expression,
\begin{equation}\label{fqt6}
\Theta_{\mu\nu} = g^{\alpha\beta}\frac{\delta T_{\alpha\beta}}{\delta g_{\mu\nu}} = g_{\mu\nu} \cL_m - 2T_{\mu\nu}.
\end{equation}

By contracting the indices $\mu$ and $\nu$ in Eq.~(\ref{fqt3}), we obtain the scalar $S$, given by
\begin{eqnarray}\label{fqt8}
S = S_\mu^{\ \mu} = -m^2 w_\mu w^\mu.
\end{eqnarray}
The trace of the matter energy-momentum tensor  $T$ can be obtained from the equation,
\begin{eqnarray}\label{fqt9}
&& \frac{1}{2} ( T + S ) - \kappa^2 f_T(T + \Theta)
= -2\kappa^2 f + 6\kappa^2 f_Q w_\mu w^\mu +\nn
&& \lambda( R - 6w_\mu w^\mu + 12 \nabla_\mu w^\mu ) + 6w^\mu \nabla_\mu \lambda + 3\Box \lambda
\end{eqnarray}

By taking the covariant divergence of Eq.(\ref{fqt3}), and using the constraint $\tilde R = 0$, we obtain for the divergence of the ordinary matter energy-momentum tensor the expression
\begin{eqnarray}\label{fqt10}
\tilde \nabla^\mu T_{\mu\nu} &=& \nabla^\mu T_{\mu\nu} - 2w^\mu T_{\mu\nu} + w_\nu T
= \frac{\kappa^2}{1+2\kappa^2f_T}\times\nn
&& \bigg[2\nabla_\nu(f_T\cL_m) - f_T\nabla_\nu T - 2T_{\mu\nu}\nabla^\mu f_T\bigg] -\nn
&& 2w^\mu T_{\mu\nu} + w_\nu T.
\end{eqnarray}

Hence, as one can see from the above equation, the ordinary  matter energy-momentum tensor is not conserved in the Weyl-type $f(Q,T)$ theory. From a physical point of view the nonconservation of the matter energy-momentum tensor can be interpreted as indicating the presence of an extra-force, acting on massive test particles, and making the motion nongeodesic. It should be noted that in the special case $f_T = 0$, the energy-momentum tensor becomes conserved, as one can be easily seem from Eq.~\eqref{fqt3}. In this case the term in the action generated by the non-minimal coupling between matter and geometry does vanish, and the theory reduces to $f(Q)$ coincident gravity \cite{beltran2018}.

There is a special case where the denominator of Eq.~(\ref{fqt10}) vanishes identically,  and therefore the field equations  Eq.~(\eqref{fqt3}) are not valid. This particular case corresponds to $1+2\kappa ^2f_T=0$, giving immediately $f_T = - 1/2\kappa^2$, and $f(Q,T) = - T/2\kappa^2 + C(Q)$, where $C(Q)$ is an arbitrary function depending only on the scalar nometricity. For this form of $f(Q,T)$  the gravitational field equations Eqs.~(\ref{fqt3}) become,
\begin{align}\label{fqt11}
	&\frac{1}{2}\l(S_{\mu\nu} + g_{\mu\nu} \cL_m\r) \nn
	&= -\frac{1}{2}\kappa^2C(Q)g_{\mu\nu} + \frac{T}{4\kappa^2} g_{\mu\nu} - 6\kappa^2C_Qw_\mu w_\nu +\nn
	& \quad \lambda\l(R_{\mu\nu}-6w_\mu w_\nu + 3g_{\mu\nu} \nabla_\rho w^\rho\r) +\nn
	& \quad 3g_{\mu\nu} w^\rho\nabla_\rho \lambda- 6w_{(\mu}\nabla_{\nu)}\lambda+g_{\mu\nu}\Box\lambda - \nabla_\mu\nabla_\nu \lambda,
\end{align}
where $C_Q = \partial C(Q)/ \partial Q$. The conservation equation of the matter energy-momentum tensor reduces to the simple form
\begin{eqnarray}\label{fqt12}
	\nabla_\nu \left(T - 2\cL_m\right) = 0.
\end{eqnarray}

\section{Geodesic deviation equation in $f(Q,T)$ theory}\label{sect2}

As a first step in our further investigations of the geometric and physical properties of the Weyl type $f(Q,T)$ gravity, in the present Section we derive the geodesic deviation equation in this theory, and obtain its form by taking into account the presence of the extra-force generated by the nonconservation of the energy-momentum tensor.

\subsection{The extra force in Weyl-type $f(Q,T)$ theory}

As we have already mentioned, due to the presence of nonmetricity, in Weyl geometry the length of a vector is no longer preserved under parallel transport. However, we define the four-velocity, as usual, according to the expression  $u^\mu = d x^\mu/d\xi$,  where $\xi$ is the affine parameter, which forms a tangent bundle of congruences of timelike curves. But in a Weyl geometry we need to normalize the four-velocity according to \cite{iosifidis2018},
\begin{eqnarray}\label{gde1}
u_\mu u^\mu = g_{\mu\nu} u^\mu u^\nu = -\ell^2,\quad \ell = \ell(x^\alpha),
\end{eqnarray}
where $\ell(x^\alpha)$ is an arbitrary  function of space and time coordinates. With this normalization, even if we set the proper time as $d\tau^2 = -g_{\mu\nu} dx^\mu dx^\nu$, $\xi$ does not necessarily coincide with $\tau$ \cite{iosifidis2018}, and by chain-rule of differentiation we obtain the relation $d\tau/d\xi = \pm \ell$.

The presence of nonmetricity, as well as its specific properties related to the background spacetime strongly affects the nature of the hypersurfaces orthogonal to the timelike $u^\mu$ velocity field. Therefore, in Weyl geometry, a generalized projection tensor operator needs to be introduced by the definition \cite{iosifidis2018}
\begin{equation}\label{gde2}
h_{\mu\nu} = g_{\mu\nu} + \frac{1}{\ell^2} u_\mu u_\nu.
\end{equation}
The generalized projection operator has the usual properties  $h_{\mu\nu} = h_{\nu\mu}$, and $h_{\mu\nu} h^{\mu\nu} = 3$, respectively. The mixed tensor form of the projection operator can be obtained by raising one of its indices, and thus
\begin{equation}\label{gde3}
h_{\mu\lambda} h^{\lambda\nu} = h_\mu^{\ \nu} = \delta_\mu^{\ \nu} + \frac{1}{\ell^2} u_\mu u^\nu.
\end{equation}

For mathematical convenience, here onwards we denote the temporal derivative by  ${\ }'$, to indicate covariant differentiation with respect to $\xi$, and the spatial derivative $D_\mu$ for a generalized tensor in Weyl geometry, respectively, as follows \cite{iosifidis2018},
\begin{eqnarray}\label{gde4}
T_{\alpha_1\dots\alpha_m}^{\beta_1\dots\beta_n\prime} &=& u^\mu\tn_\mu T_{\alpha_1\dots\alpha_m}^{\beta_1\dots\beta_n},
\end{eqnarray}
\begin{eqnarray}\label{gde5}
D_\mu T_{\alpha_1\dots\alpha_m}^{\beta_1\dots\beta_n} &=& h_\mu^{\ \lambda} h_{\alpha_1}^{\ \gamma_1}\dots h_{\alpha_m}^{\ \gamma_m}h_{\delta_1}^{\ \beta_1}\dots h_{\delta_n}^{\ \beta_n} \tn_\lambda^{\ } T_{\gamma_1\dots\gamma_m}^{\delta_1\dots\delta_n}.\nn
\end{eqnarray}

Due to the presence  of nonmetricity in Weyl geometry there exist two types of four-acceleration, denoted by $A^\mu$ and $a_\mu$, respectively, and defined according to
\begin{eqnarray}\label{gde6}
A^\mu &=& u^{\mu\prime} = u^\nu \tn_\nu u^\mu = u^\nu \nabla_\nu u^\mu - \ell^2 w^\mu - 2w_\nu u^\nu u^\mu,\nn
\end{eqnarray}
and
\begin{eqnarray}
a_\mu &=& u_\mu' = u^\nu \tn_\nu u_\mu = u^\nu \nabla_\nu u_\mu - \ell^2 w_\mu,
\end{eqnarray}
respectively. The two accelerations are related by the important equation,
\begin{eqnarray}\label{gde7}
	A^\mu &=& a^\mu + Q^{\nu\lambda\mu}u_\nu u_\lambda.
\end{eqnarray}

From these two types of acceleration we can obtain some pure geometrical relations. Multiplying the two acceleration vectors by $u_\mu$, we obtain
\begin{eqnarray}\label{gde8}
A^\mu u_\mu &=& -\frac{1}{2} (\ell^2)' + \frac{1}{2} Q_{\mu\nu\lambda} u^\mu u^\nu u^\lambda,\\
a^\mu u_\mu &=& -\frac{1}{2} (\ell^2)' - \frac{1}{2} Q_{\mu\nu\lambda} u^\mu u^\nu u^\lambda.
\end{eqnarray}
For the sake of clarity we mention that $(\ell^2)^\prime = u^\mu \tn_\mu (\ell^2)$.

With the above definitions, the geodesic equation in Weyl geometry takes the following form, where generally the extra force $f^\mu$ gives the supplementary acceleration induced by the nonminimal curvature-matter coupling,
\begin{eqnarray}\label{gde9}
u^{\mu\prime} = u^\nu \tn_\nu u^\mu = \frac{d^2 x^\mu}{d\xi^2} + \tg_{\nu\lambda}^\mu u^\nu u^\lambda = f^\mu.
\end{eqnarray}

The energy-momentum tensor  of a perfect fluid can be defined in a Weyl spacetime according to \cite{iosifidis2018}
\begin{equation}\label{gde10}
T_{\mu\nu} = \frac{(p+\rho)}{\ell^2} u_\mu u_\nu + p g_{\mu\nu}.
\end{equation}
The four-velocity $u^\mu$ of the perfect fluid must be $\xi$-parameterized according to the definition of velocity in Weyl geometry, and it obeys the normalization condition (\ref{gde1}). The details of the derivation of the generalization of the perfect fluid model in Weyl geometry can be found in Appendix~\ref{app3}.

Now multiplying with the projection operator the covariant divergence of the energy-momentum tensor in the Weyl-type $f(Q,T)$ theory, we obtain the important correspondence between the matter energy-momentum tensor $T_{\mu\nu}$ and the four-acceleration $A^\mu$ as
\begin{eqnarray}\label{gde11}
h^{\rho\nu} \tn^\mu T_{\mu\nu} &=& \frac{p+\rho}{\ell^2} (A^\rho - Q^{\mu\nu\rho} u_\mu u_\nu) -\nn
&& \frac{p+\rho}{2\ell^4} \bigg[(\ell^2)' + Q^{\mu\nu\lambda} u_\mu u_\nu u_\lambda\bigg] u^\rho +\nn
&& h^{\rho\nu} \tn_\nu p + h^{\nu\rho} Q^\mu_{\ \mu\nu} p.
\end{eqnarray}
With the use of  Eqs.(\ref{gde6}) and (\ref{gde9}), from Eq.~(\ref{gde11}) we can obtain the expression of the extra force in Weyl-type $f(Q,T)$ gravity as follows (for the calculational details see Appendix~\ref{app4}),
\begin{eqnarray}\label{gde12}
\hspace{-0.5cm}f^\rho &=& u^{\mu\prime}
= \frac{\ell^2}{p+\rho} h^{\rho\nu} \tn^\mu ( T_{ \mu \nu } - p g_{ \mu \nu } ) +\nn
\hspace{-0.5cm}&& \frac{1}{2\ell^2} \bigg[(\ell^2)' + Q^{\mu\nu\lambda} u_\mu u_\nu u_\lambda \bigg] u^\rho + Q^{\mu\nu\rho} u_\mu u_\nu.
\end{eqnarray}

By substituting the expression of the nonmetricity tensor in the above equation, we obtain for the extra force the final expression
\begin{eqnarray}\label{gde13}
f^\rho &=& \frac{\ell^2}{p+\rho} h^{\rho\nu} \tn^\mu ( T_{ \mu \nu } - p g_{ \mu \nu } ) + \frac{(\ell^2)'}{2\ell^2} u^\rho - w_\mu u^\mu u^\rho.\nn
\end{eqnarray}

\subsection{The geodesic deviation equation}

Consider now a one-parameter congruence of curves $x^\mu(\xi;\sigma)$ satisfying the geodesic equation for the parameter $\xi$ in Weyl geometry for each $\sigma = \sigma_0 = {\rm constant}$. We introduce the four-vectors
\begin{eqnarray}\label{gde14}
U^\mu = \frac{\partial x^\mu(\xi;\sigma)}{\partial \sigma}, \quad \eta^\mu = U^\mu \delta \sigma.
\end{eqnarray}

The second temporal derivative of $U^\mu$ can be expressed as
\begin{eqnarray}\label{gde15}
U^{\mu\prime\prime} &=& u^\nu \tn_\nu (u^\alpha \tn_\alpha U^\mu)
= u^\nu \tn_\nu (U^\alpha \tn_\alpha u^\mu)\nn
&=& (\tn_\nu \tn_\alpha u^\mu) U^\alpha u^\nu + (\tn_\nu U^\alpha) (\tn_\alpha u^\mu) u^\nu.
\end{eqnarray}

Using the definition of curvature tensor, we have
\begin{eqnarray}\label{gde16}
U^{\mu\prime\prime} &=& ( - \tilde R^\mu_{\ \beta\alpha\nu} u^\beta + \tn_\alpha \tn_\nu u^\mu) U^\alpha u^\nu + U^\nu(\tn_\nu u^\alpha) (\tn_\alpha u^\mu)\nn
&=& - \tilde R^\mu_{\ \beta\alpha\nu} U^\alpha u^\beta u^\nu + U^\alpha \tn_\alpha(u^\nu \tn_\nu u^\mu).
\end{eqnarray}

Hence we obtain the geodesic deviation equation in Weyl geometry in the presence of an extra force as,
\begin{eqnarray}\label{gde17}
U^{\mu\prime\prime} &=& - \tilde R^\mu_{\ \nu\alpha\beta} U^\alpha u^\beta u^\nu + U^\alpha \tn_\alpha f^\mu.
\end{eqnarray}

By multiplying with $\delta\sigma$  both sides of the above equation gives
\begin{equation}\label{gde18}
\eta^{\mu\prime\prime} = - \tilde R^\mu_{\ \nu\alpha\beta} \eta^\alpha u^\beta u^\nu + \eta^\alpha\tn_\alpha f^\mu.
\end{equation}

Eq.~(\ref{gde18}) represents the geodesic deviation equation in Weyl geometry in the presence of an extra force generated by the coupling between nonmetricity and matter, as follows from the mathematical and physical structure of $f(Q,T)$ gravity. In the special case when $f_T=0$, the energy-momentum tensor becomes conserved, and Eq.~\eqref{gde18} reduces to the geodesic deviation equation in Weyl geometry, as considered in \cite{iosifidis2018}. Note that in this case the present theory becomes equivalent to the coincident gravity theory \cite{beltran2018}. In the more general case where the Weyl-vector also vanishes, the space-time becomes flat and the right hand side of Eq.~ \eqref{gde18} identically vanishes.

\section{Generalized Raychaudhuri equation in Weyl-type $f(Q,T)$ theory}\label{sect3}

In the present Section we derive the Raychaudhuri equation in the Weyl-type $f(Q,T)$ theory, and in the presence of an extra force. Our approach basically follows, and generalizes,  the similar analysis performed in \cite{iosifidis2018}, in which the presence of the extra-force has not been taken into account.

We begin our investigation by decomposing the covariant derivative of the four-velocity $u_\mu$  into its temporal and spatial components, according to \cite{iosifidis2018},
\begin{equation}\label{ray1}
\tn_\nu u_\mu = D_\nu u_\mu - \frac{1}{\ell^2}(u_\mu \xi_\nu + a_\mu u_\nu) - \frac{1}{\ell^4} (u^\lambda a_\lambda) u_\mu u_\nu,
\end{equation}
where $D_\nu u_\mu = h_\nu^{\ \beta} h_\mu^{\ \lambda} \tn_\beta u_\lambda$ and $\xi_\mu = u^\nu \tn_\mu u_{\nu} = u^\nu \nabla_\mu u_\nu - \ell^2 w_\mu$. We can see that by construction we have the relation $\xi_\mu u^\mu = a_\mu u^\mu$.

Similarly, the projected covariant derivative can be decomposed into
\begin{equation}\label{ray2}
D_\nu u_\mu = \frac{1}{3}\bigg(\theta + \frac{1}{\ell^2} a_\alpha u^\alpha\bigg) h_{\mu\nu} + \sigma_{\mu\nu} + \omega_{\mu\nu},
\end{equation}
where
\begin{equation}\label{ray3}
\theta = g^{\mu\nu}\tn_\nu u_{\mu} = D^\mu u_\mu - \frac{1}{\ell^2}a_\mu u^\mu = \nabla^\mu u_\mu - 2w_\mu u^\mu,
\end{equation}
is the "volume" scalar in Weyl geometry. We can also construct the scalar $D^\mu u_\mu = h^{\mu\nu} \tn_\nu u_\mu$.

In a Weyl geometry and in the presence of nonmetricity the shear tensor is defined according to
\begin{equation}\label{ray4}
\sigma_{\mu\nu} = D_{\langle\nu}u_{\mu\rangle} = h_{(\nu}^{\ \beta} h_{\mu)}^{\ \lambda} \tn_\beta u_\lambda - \frac{1}{3}\bigg(\theta + \frac{1}{\ell^2} a_\alpha u^\alpha \bigg) h_{\mu\nu},
\end{equation}
while the vorticity tensor is introduced based on the definition
\begin{equation}\label{ray5}
\omega_{\mu\nu} = D_{[\nu}u_{\mu]} = \frac{1}{2} (D_{\nu} u_{\mu } - D_{\mu} u_{\nu}).
\end{equation}

 The shear tensor is symmetric and trace-free due to its construction. As for the  vorticity tensor, it is naturally anti-symmetric. Thus, based on these mathematical properties,  we find that the following relations are always satisfied, $\sigma_\mu^{\ \mu} = 0 = \omega_\mu^{\ \mu}$, $\sigma_{\mu\nu} u^\nu = 0 = \omega_{\mu\nu} u^\nu$, $\sigma_{\mu\nu}h^{\mu\nu} = 0 = \omega_{\mu\nu}h^{\mu\nu}$ and $\sigma_{\mu\nu}\omega^{\mu\nu} = 0$.

From Eqs.~(\ref{ray1}) and (\ref{ray2}) we obtain the explicit expression of the covariant derivative of $u_\mu$ as
\begin{eqnarray}\label{ray6}
\tn_\nu u_\mu &=& \frac{1}{3} \bigg(\theta + \frac{1}{\ell^2} a_\alpha u^\alpha \bigg) h_{\mu\nu} + \sigma_{\mu\nu} + \omega_{\mu\nu} -\nn
&& \frac{1}{\ell^2} (u_\mu\xi_\nu + a_\mu u_\nu) - \frac{1}{\ell^4}(u^\alpha a_\alpha) u_{\mu} u_{\nu},
\end{eqnarray}
while for the covariant derivative of $u^\mu$ we find
\begin{eqnarray}\label{ray7}
\tn^\nu u^\mu &=& \frac{1}{3} \bigg(\theta + \frac{1}{\ell^2} a_\alpha u^\alpha\bigg) h^{\mu\nu}  - \frac{1}{\ell^2}(u^\mu \xi^\nu + a^\mu u^\nu) - \nonumber\\
&& \frac{1}{\ell^4}(u^\alpha a_\alpha) u^\mu u^\nu + Q^{\nu\mu\alpha} u_\alpha + \sigma^{\mu\nu} + \omega^{\mu\nu}.\nn
\end{eqnarray}

Raychaudhuri's equation is a purely geometrical relation, and it immediately follows from a set of fundamental geometric relations, known as Ricci's identities. In a Weyl spacetime and in the presence of nonmetricity, the definition of the curvature tensor is,
\begin{equation}\label{ray8}
(\tn_\mu\tn_\nu - \tn_\nu\tn_\mu)u_\lambda = \tilde R^\beta_{\ \lambda\nu\mu}u_\beta.
\end{equation}
We multiply both sides of Eq.(\ref{ray8}) with $g^{\lambda\nu}u^\mu$, thus obtaining
\begin{eqnarray}\label{ray9}
g^{\lambda\nu}u^\mu \tn_\mu\tn_\nu u_\lambda - g^{\lambda\nu}u^\mu  \tn_\nu\tn_\mu u_\lambda &=& - \tilde R_{\beta\lambda\mu\nu}u^\beta u^\mu g^{\lambda\nu}.\nn
\end{eqnarray}

The first term on the left hand side of Eq.~(\ref{ray9}) can be evaluated as
\begin{eqnarray}\label{ray10}
&& g^{\lambda\nu}u^\mu \tn_\mu\tn_\nu u_\lambda
= \theta' - \frac{1}{3}\bigg(\theta + \frac{1}{\ell^2} a_\alpha u^\alpha\bigg) u^\mu Q_\mu -\nonumber\\
&& u^\mu \sigma^{\nu\lambda} Q_{\mu\nu\lambda} -
 \frac{1}{3\ell^2} \bigg( \theta - \frac{2}{\ell^2} a_\alpha u^\alpha \bigg) u^\mu u^\nu u^\lambda Q_{\mu\nu\lambda} +\nonumber\\
&& \frac{1}{\ell^2} (\xi^\lambda + a^\lambda)u^\mu u^\nu Q_{\mu\nu\lambda},
\end{eqnarray}
while the second term on the left hand side of Eq.~(\ref{ray9}) evaluates to
\begin{eqnarray}\label{ray11}
&& g^{\lambda\nu}u^\mu \tn_\nu\tn_\mu u_\lambda
= -\frac{1}{3} \theta^2 + \tn^\mu a_\mu - \frac{2}{3\ell^2} \theta a_\alpha u^\alpha + \nonumber\\
&&\frac{2}{3\ell^4} (a_\alpha u^\alpha)^2 -
 \frac{1}{3} \bigg(\theta + \frac{1}{\ell^2} a_\alpha u^\alpha\bigg) u^\mu_{\ } Q_{\ \nu\mu}^\nu + \frac{2}{\ell^2} a_\alpha \xi^\alpha - \nn
&&\frac{1}{3\ell^2} \bigg(\theta - \frac{2}{\ell^2} a_\alpha u^\alpha\bigg) u^\mu u^\nu u^\lambda Q_{\mu\nu\lambda} +\nn
&& \frac{1}{\ell^2} (u^\mu \xi^\nu + a^\mu u^\nu) u^\lambda Q_{\mu\nu\lambda} - \sigma_{\mu\nu} \sigma^{\mu\nu} + \omega_{\mu\nu} \omega^{\mu\nu} -\nn
 &&(\sigma^{\mu\nu} + \omega^{\mu\nu}) u^\lambda Q_{\mu\nu\lambda}.
\end{eqnarray}

Hence in Weyl geometry the left hand side of Eq.~(\ref{ray9}) finally reads,
\begin{eqnarray}\label{ray12}
&&g^{\lambda\nu}u^\mu \tn_\mu\tn_\nu u_\lambda - g^{\lambda\nu}u^\mu  \tn_\nu\tn_\mu u_\lambda \nn
&=& \theta' + \frac{1}{3} \theta^2 - \tn^\mu a_\mu + \frac{2\theta}{3\ell^2} a_\alpha u^\alpha - \frac{2}{3\ell^4} (a_\alpha u^\alpha)^2 -\nn
&& \frac{2}{\ell^2} a_\alpha \xi^\alpha + \sigma_{\mu\nu} \sigma^{\mu\nu} - \omega_{\mu\nu}\omega^{\mu\nu} + (\sigma^{\mu\nu} + w^{\mu\nu}) u^\lambda Q_{\mu\nu\lambda} + \nn
&& \frac{1}{3} \bigg(\theta + \frac{1}{\ell^2} a_\alpha u^\alpha\bigg) u^\mu Q^\nu_{\ \nu\mu} - \frac{1}{3} \bigg(\theta + \frac{1}{\ell^2} a_\alpha u^\alpha \bigg) u^\mu Q_\mu -\nn
&& u^\mu \sigma^{\nu\lambda} Q_{\mu\nu\lambda} + \frac{1}{\ell^2} (u^\mu u^\nu a^\lambda - a^\mu u^\nu u^\lambda) Q_{\mu\nu\lambda}\nn
&=& \bigg[\theta - \frac{(\ell^2)'}{2\ell^2} - w_\mu u^\mu\bigg]' + \frac{1}{3} \bigg[\theta - \frac{(\ell^2)'}{2\ell^2} - w_\mu u^\mu\bigg]^2 +\nn
&& \sigma_{\mu\nu} \sigma^{\mu\nu} - \omega_{\mu\nu}\omega^{\mu\nu} - \nabla^\mu f_\mu+2w^\mu f_\mu+\frac{1}{\ell^2}f^\mu\nabla_\mu \ell^2 + \nn
&& \bigg[\frac{(\ell^2)'}{2\ell^2} - w_\mu u^\mu\bigg]' - \bigg[\frac{
	(\ell^2)'}{2\ell^2} - w_\mu u^\mu\bigg]^2.
\end{eqnarray}

The right hand side of the above equation can be evaluated as
\begin{eqnarray}\label{ray13}
\hspace{-0.6cm}\tilde R_{\beta\lambda\mu\nu} u^\beta u^\mu g^{\lambda\nu} &=& R_{\mu\nu} u^\mu u^\nu + u^\mu u^\nu \tn_\nu Q^\alpha_{\ \alpha\mu} -\nn
\hspace{-0.6cm}&& u^\mu u^\nu \tn^\alpha Q_{\mu\nu\alpha} - Q_\mu^{\ \alpha\beta} Q_{\beta\alpha\nu} u^\mu u^\nu \nn
\hspace{-0.6cm}&=& R_{\mu\nu} u^\mu u^\nu + 2u^\mu u^\nu \nabla_\mu w_\nu +\nn
\hspace{-0.6cm}&& 2 (w_\mu u^\mu)^2 - \ell^2 \nabla^\mu w_\mu + 2\ell^2 w_\mu w^\mu.
\end{eqnarray}

With the use of Eqs.(\ref{ray12}) and (\ref{ray13}) we finally obtain the Raychaudhuri equation in Weyl geometry, and in the presence of an extra force, as,
\begin{eqnarray}\label{ray14}
&& \bigg[\theta - 2 w_\mu u^\mu\bigg]'= -\frac{1}{3} \bigg[\theta - \frac{(\ell^2)'}{2\ell^2} - w_\mu u^\mu\bigg]^2 +\nn
&& \bigg[\frac{(\ell^2)'}{2\ell^2} - w_\mu u^\mu\bigg]^2 - R_{\mu\nu} u^\mu u^\nu -\nn
&& \sigma_{\mu\nu} \sigma^{\mu\nu} + \omega_{\mu\nu}\omega^{\mu\nu} + \nabla^\mu f_\mu-\f{1}{\ell^2}f^\mu\nabla_\mu\ell^2-2w_\mu f^\mu-\nn
&& 2 (w_\mu u^\mu)^2  - 2u^\mu u^\nu \nabla_\mu w_\nu + \ell^2 \nabla^\mu w_\mu - 2\ell^2 w_\mu w^\mu.\nn
\end{eqnarray}

It is worth mentioning again that the effect of the non-minimal matter-geometry coupling $f(Q,T)$ will enter into the Raychaudhury equation (\ref{ray14}) through the expression of the extra force, given by Eq.~\eqref{gde13}. As we have already mentioned earlier, in the case of minimal matter-geometry coupling $f_T=0$, the above equation reduces to the generalized Raychaudhuri equation in the coincidence gravity theory \cite{beltran2018}. As is well-known, the coincidence gravity is a generalization of the symmetric teleparallel gravity \cite{beltran2018}. In this sense, Eq.~(\ref{ray14}) with $f_T=0$ could be considered as the Raychaudhuri equation of the generalized symmetric teleparallel equivalent to GR.

 The special case with zero extra force of the Raychaudhuri equation has been obtained and discussed in \cite{iosifidis2018}. The first three lines in Eq.~(\ref{ray14}) have analogue forms to the similar terms in modified gravity theories with geometry-matter coupling formulated in Riemann geometry \cite{harko2012}. The Raychaudhuri and optical equations for null geodesic congruences with torsion were investigated in \cite{Speziale}.

\section{Weak field approximation,  Newtonian and Post-Newtonian li,its, tidal force and Roche radius in Weyl-type $f(Q,T)$ gravity theory}\label{sect4}

In the present Section, based on the previous mathematical results,  we will consider the weak field limit of the Weyl-type $f(Q,T)$ gravity theory. The generalized Poisson equation, as well as the expressions of the tidal force tensor are derived. As an astrophysical application of the obtained results we consider the modifications to the Roche limit that are induced by the assumption of the presence of nonmetricity in the space-time geometry.

\subsection{The weak field approximation}

If one considers the physical situation in which the motion of the test particles is slow, and that the gravitational field intensity created by the material particles is comparably weak, one could easily compare generalized metric theories of gravity with each other, with the experimental observations, as well as with Newtonian gravity. In this case, the first order approximation is adequately accurate to compare the theoretical predictions of the gravitational theories with past, present and future Solar System observations. Generally, this kind of approximation of gravitational theories, the so-called post-Newtonian limit, is valid within the near-zone of the system, corresponding to a spherical region with size smaller than one gravitational wavelength.

As a first step in our investigation of the physical properties of the $f(Q,T)$ gravity theory, in the following we investigate the linear approximation of the metric field, or the weak field approximation, by assuming the decomposition of the metric tensor as
\begin{eqnarray}\label{til1}
g_{\mu\nu} &=& \eta_{\mu\nu} + H_{\mu\nu}, \quad \l| H_{\mu\nu}\r | \ll 1,
\end{eqnarray}
where $\eta_{\mu\nu} = {\rm diag}(-1,1,1,1)$ is the metric tensor in the Minkowski spacetime. In Weyl geometry, by keeping only the first order of $H_{\mu\nu}$, the connection and curvature tensor can be respectively formed from their Riemann part \cite{gron}, to which we add  the non-Riemannian (Weyl) components of the connection
\begin{eqnarray}\label{til2}
\tg_{\lambda\mu\nu} &=& \frac{1}{2} (\partial_\nu H_{\mu\lambda}+\partial_\mu H_{\nu\lambda} - \partial_\lambda H_{\mu\nu}) +\nn
&& g_{\mu\nu} w_\lambda - g_{\lambda\mu} w_\nu - g_{\lambda\nu} w_\mu,
\end{eqnarray}
and of the curvature tensor,
\begin{eqnarray}\label{til3}
&& \tilde R_{\alpha\mu\beta\nu} \nn
&=& \frac{1}{2} (\partial_\beta\partial_\mu H_{\nu\alpha} + \partial_\nu\partial_\alpha H_{\mu\beta} - \partial_\beta\partial_\alpha H_{\mu\nu} - \partial_\nu\partial_\mu H_{\alpha\beta}) \nn
&& + 2\nabla_\beta w_{[\alpha}g_{\mu]\nu} + 2\nabla_\nu w_{[\mu}g_{\alpha]\beta} + 2w_\beta w_{[\alpha}g_{\mu]\nu} +\nn
&& 2w_\nu w_{[\mu}g_{\alpha]\beta} - 2 w^\lambda w_\lambda g_{\beta[\alpha}g_{\mu]\nu} + g_{\alpha\mu} W_{\beta\nu},
\end{eqnarray}
respectively. In the following the {\it Latin letters denote the spatial components} ($i=1,2,3$) of the tensors, and we will use {\it Greek letters to indicate both spatial and temporal components} ($\mu = 0,1,2,3$).

By assuming the metric is diagonal and time-independent, indicating that $H_{\mu\nu}$ is also a static field, one can obtain the explicit weak-field expressions of the Christoffel part in curvature tensor as \cite{gron, hans},
\begin{eqnarray}\label{til5}
\Gamma^\mu_{\ 00} = -\frac{1}{2} g^{\mu\nu} \partial_\nu g_{00} \approx - \frac{1}{2} \eta^{\mu\mu} \partial_\mu H_{00},
\end{eqnarray}
\begin{eqnarray}\label{til6}
\Gamma^0_{\ 00} = - \frac{1}{2} \eta^{00} \partial_0 H_{00} = 0,\quad \Gamma^i_{\ 00} = - \frac{1}{2} \eta^{ii} \partial_i H_{00}.
\end{eqnarray}
From Eq.~(\ref{til3}) we obtain the component of $R^\gamma_{~ 0 \beta 0}$,
\begin{eqnarray}\label{til24}
	R^\gamma_{~ 0 \beta 0} &=& -\frac{1}{2} \partial_\beta \partial_\gamma H_{00},
\end{eqnarray}
and
\begin{eqnarray}\label{til11}
R_{\mu\nu} &=& \frac{1}{2} \times\nn
&& \l( \partial_\mu \partial_\alpha H^\alpha_{\ \nu} + \partial_\nu \partial_\alpha H^\alpha_{\ \mu} - \partial_\nu \partial_\mu H^\alpha_{\ \alpha} - \partial^\alpha \partial_\alpha H_{\mu\nu} \r).\nn
\end{eqnarray}
For the details of the calculations of the Riemann and Ricci tensors in the weak field approximation see Appendix~\ref{app7}.

\subsection{Post-Newtonian analysis}

In this Section we will discuss in detail the Newtonian limit of $f(Q,T)$ gravity, by assuming that the perfect fluid filling the space-time is nonrelativistic. First of all, we will chose a reference frame where the motion of the fluid is static. Moreover, we assume that matter is in the form or pressureless dust with the property  $\rho \gg p$, and that the velocity of particles is small as compared to the speed of light.

Hence we can approximate the four-velocity of the fluid as $u^\mu = (\ell,u^i)$, and $u_\mu = (-\ell,u_i)$, respectively, and we keep only the first order terms in $u^i$. In the framework of these approximations  the matter energy-momentum tensor $T_{\mu\nu}$ reads,
\be\label{em1}
T_{\mu\nu} = \frac{\rho}{\ell^2} u_\mu u_\nu.
\ee

Assuming that $\lambda=\lambda_0+\delta\lambda$ and $\ell=1+\delta\ell$, where $\delta\lambda$ and $\delta\ell$ are perturbation variables, from Eq.(\ref{weyl5}) and Eq.(\ref{fqt2}), we obtain the first-order perturbation of $\delta \lambda$ and $\delta \ell$ as
\begin{eqnarray}
\delta x^\mu \partial_\mu \lambda = \delta \lambda &=& - (2\kappa^2 f_Q + 2\lambda_0)w_\mu \delta x^\mu\nn
&=& - (2\kappa^2 f_Q + 2\lambda_0)\delta \ell,
\end{eqnarray}
Using the above equation together with equation \eqref{em1}, one finds that the only non-vanishing component of the energ-momentum tensor is $T^{0}_{~0}=\rho$.

It should be noted that in the Newtonian limit, one decompose the tensor $H_{\mu\nu}$ as
\begin{equation}\label{til10}
\Phi = - \frac{1}{2} H_{00},\quad\Psi = - \frac{1}{2} H_{ii}.
\end{equation}
where $\Phi$ and $\Psi$ are Newtonian potentials. Also we assume that the Weyl vector can be written as $\omega_\mu=(0,\omega_i)$ where $\omega_i\sim\mathcal{O}(\epsilon)$.

From these definitions, one can see that the tensor $S_{\mu\nu}$ in \eqref{fqt4} becomes second order in perturbation variables and does not contribute to the Newtonian limit of the theory.

 Considering the metric field equations \eqref{fqt3} at background level, it follows that the background value of the Lagrange multiplier $\lambda_0$ should be constant, and also the condition $f^{(0)}=0$, where by $(0)$ we mean the background value, must hold.

By expanding the field equation \eqref{fqt3} up to first order in the perturbation parameters, and by using the background constraints derived above, one can obtain the $(00)$ and $(ii)$ components of the field equations as
\begin{align}
&2\lambda_0\Delta\Phi-2\Delta\delta\lambda=6\lambda_0\nabla_i\omega_i+(1+\kappa^2f_T^{(0)})\rho,\label{com00}\\
&\lambda_0(\Delta\Phi-4\Delta\Psi)-2\Delta\delta\lambda=9\lambda_0\nabla_i\omega_i-\f32\kappa^2 f_T^{(0)}\rho.\label{comii}
\end{align}

Also, from the off-diagonal components of the metric field equation \eqref{fqt3}, we obtain,
\begin{align}
\delta\lambda=\lambda_0(\Psi-\Phi).\label{ofdiag}
\end{align}
In order to close the system of dynamical equations, one should also consider the constraint equation $\tilde{R}=0$, and also the vector field equation \eqref{fqt2} to first order in perturbations. The constraint equation becomes
\begin{align}
\Delta\Phi-2\Delta\Psi=3\nabla_i\omega_i,\label{const}
\end{align}
and the divergence of the vector field equation becomes
\begin{align}
6\lambda_0(\Delta\Phi&-\Delta\Psi)\nonumber\\&=\f13(\Delta\Phi-2\Delta\Psi)\left(m^2+12\lambda_0+12\kappa^2f_Q^{(0)}\right).\label{vec1}
\end{align}

Now, by substituting $\nabla_i\omega_i$ and $\Delta\delta\lambda$ from equations \eqref{const} and \eqref{ofdiag} in the $(ii)$ component of the metric field equation \eqref{comii}, one can see that $f_T^{(0)}\approx0$. Now, solving the remaining equations \eqref{com00} and \eqref{vec1} for the Newtonian potentials, we find
\begin{align}\label{nl6}
\Delta \Phi &= \f12G_{eff}\rho,
\end{align}
\begin{align}
\Delta \Psi &= \f12\gamma G_{eff}\rho,
\end{align}
where we have defined the generalized Newtonian gravitational constant $G_{eff}$ as
\begin{align}
G_{eff}=2\left(\f{1}{3\lambda_0}+\f{1}{m^2+12\kappa^2f_Q^{(0)}}\right),
\end{align}
and the PPN $\gamma$ parameter as
\begin{align}
\gamma=\f{m^2+12\kappa^2f_Q^{(0)}-6\lambda_0}{2(m^2+12\kappa^2f_Q^{(0)}+3\lambda_0)}.
\end{align}

One can see from its definition that the generalized Newton gravitational constant depends on the derivative of the function $f(Q,T)$ with respect to the trace of the energy-momentum tensor $T$, on the mass of the Weyl vector field, and on the background value of the Lagrange multiplier.

It is worth mentioning that the value $\gamma=1$, which corresponds to the GR result, and which is confirmed by the observations at the Solar System level, occurs when $f_Q^{(0)}=-\lambda_0\kappa^2-m^2/12\kappa^2$. In this case we obtain $\lambda_0=1/2G$, where $G$ is the Newtonian constant.

The present analysis does show that the effects of the matter geometry couplings $f(Q,T)$ do appear already in the first-order perturbation of the theory. This is in fact different from the generalized teleparallel theory, where the first-order perturbation analysis does not reveal the extra degrees of freedom, and one should take into account higher order perturbation analysis \cite{higher}. However, in order to explore the detailed dependence of the function $f(Q,T)$ on the trace of the energy-momentum tensor $T$, one should consider higher order perturbation analysis of the model.

The extra force \eqref{gde13}, can be decomposed into an energy-momentum related component, and a geometry related component,
\begin{eqnarray}\label{nl17}
	f^\rho &\approx& \frac{h^{\rho\nu} \tn^\mu T_{ \mu \nu }}{\rho} - w_\mu u^\mu u^\rho + \frac{(\ell^2)'}{2\ell^2} u^\rho\nn
	&=& F^\rho - w_\mu u^\mu u^\rho + \frac{(\ell^2)'}{2\ell^2} u^\rho,
\end{eqnarray}
where we have defined the tensor
\be
F^\rho\equiv\frac{h^{\rho\nu} \tn^\mu T_{ \mu \nu }}{\rho},
\ee
 which can be written explicitly with the use of Eq.~(\ref{fqt10}) as,
\begin{eqnarray}\label{nl301}
	F^\rho &=& - \frac{2\ell^2h^{\rho\nu}w^\mu T_{\mu\nu}}{\rho} + \frac{\ell^2h^{\rho\nu}w_\nu T}{\rho} - \frac{\kappa^2 \ell^2h^{\rho\nu}}{(1+2\kappa^2f_T)\rho}\times\nn
	&& \bigg[2\nabla_\nu(\rho f_T) + f_T\nabla_\nu T + 2T_{\mu\nu}\nabla^\mu f_T\bigg],
\end{eqnarray}
where $u^\mu F_\mu=0$ should be always fulfilled. Using the definition of the energy-momentum tensor and also expanding the above expression to first order in perturbation variables, one obtains
\begin{align}\label{mm}
F^\rho=-\ell^2\omega^\rho+\ell u^\rho \omega_0-\frac{2\kappa^2\ell^2}{1+2\kappa^2 f_T}h^{\rho\nu}\nabla_\nu f_T,
\end{align}
where we have assumed that the energy density varies slowly in this limit. At first order in perturbations one obtains, $F^0 = 0$. Also note the second term in Eq.~\eqref{mm} is non-vanishing only in the case of $\rho=0$. Since in the following calculations we will use only the $i$ components, we will omit this term. In the particular case of $1 + 2\kappa^2 f_T = 0$, $f_T=-1/2\kappa ^2$, we find $\lim _{f_T\rightarrow -1/2\kappa ^2}\nabla_\nu f_T/\left(1+2\kappa^2f_T\right)=0$, and thus we still obtain for the extra-force the non-trivial expression $F^\rho=-\ell^2\omega^\rho+\ell u^\rho \omega_0$.

The equation describing a world line in Weyl geometry with extra force reads,
\begin{align}\label{nl29}
	u^\nu \partial_\nu u^\mu &+ \Gamma^\mu_{\ \nu\sigma} u^\sigma u^\nu\nn&=F^\mu + w_\nu u^\nu u^\mu -w^\mu u_\nu u^\nu+ \frac{(\ell^2)'}{2\ell^2} u^\mu.
\end{align}
By adopting the linear approximation and the Newtonian limit, the spatial component in Eq.~(\ref{nl29}) becomes
\begin{eqnarray}\label{nl11}
	\frac{d^2x^i}{d\tau^2} &=& -\ell \partial_i \Phi + \ell w^i + \frac{F^i}{\ell}+\frac{(\ell^2)'}{2\ell^2} u^i.
\end{eqnarray}

This equation represents the generalization of the Newtonian equation of motion in the Weyl-type $f(Q,T)$ gravity theory.

\subsection{Tidal forces in Weyl type $f(Q,T)$ gravity theory}

In a Weyl geometries, the properties of parallel transport of a vector along a geodesic line are preserved with the important exception of the magnitude of a vector changing after transport. Similarly to Riemannian geometry, in the Weyl geometry one can always find a set of tangent spaces generated by the four velocity $u^\mu$ for every point along a world line, where their axes (with the same index) remain parallel, under parallel transport along the world line. It should be noted that these four velocities are not necessarily normalized to one in every tangent space. We have already defined $u_\mu u^\mu = -\ell^2$, and in this case we will not take $\ell$ to one. Here onward we will consider geodesic reference frame in which all connection components vanish, indicating that $\tg^\lambda_{\ \mu\nu}=0$ point wisely.

We at first go back to Eq.~(\ref{til3}) to evaluate the spatial total curvature tensor $\tilde R_{i 0 j 0}$, which is of significant important in the following part. Hence we obtain
\begin{eqnarray}\label{nl10}
	\tilde R_{i 0 j 0}
	&=& R_{i 0 j 0} + \nabla_j w_i g_{0 0} + \nabla_0 w_0 g_{ij} +\nn
	&& w_j w_i g_{0 0} + w_0^2 g_{ij} - w^2 g_{j i}g_{0 0}\nn
	&\approx& R_{i 0 j 0} + \eta_{0 0}\partial_j w_i
	= R_{i 0 j 0} - \partial_j w_i,
\end{eqnarray}
where $\Gamma^\lambda_{\ \mu\nu} = -g_{\mu\nu} w^\lambda + \delta^\lambda_\mu w_\nu + \delta^\lambda_\nu w_\mu$ is obtained from the condition of zero total connection. Raising the index $i$, we have
\begin{eqnarray}\label{til34}
\tilde R^i_{\ 0j0} = R^i_{\ 0 j 0} - \partial_j w^i .
\end{eqnarray}

Hence the geodesic deviation equation Eq.~(\ref{gde18}) will become,
\begin{eqnarray}\label{nll1}
	u^\alpha \partial_\alpha(u^\beta\partial_\beta \eta^\mu) &=& - \tilde R^\mu_{\ \nu\alpha\beta} \eta^\alpha u^\beta u^\nu + \eta^\alpha \partial_\alpha f^\mu
\end{eqnarray}
If we consider the linear approximation, the Newtonian limit, and the zero total connection for this system, and we also use $\eta^0 = 0$ to indicate that the accelerations of the particles are compared at equal times, and $f^0 = 0$, respectively, indicating that the thermodynamic parameters of the matter do not depend on time, we obtain
\begin{eqnarray}
	\ell^2 \frac{d^2 \eta^i}{dt^2} &=& - \ell^2 \tilde R^i_{\ 0 j 0} \eta^j + \eta^j \partial_j F^i.
\end{eqnarray}
Now we write the explicit expression of the curvature tensor in Weyl geometry by using Eq.~(\ref{til34}) as,
\begin{eqnarray}\label{til33}
	\frac{d^2 \eta^i}{dt^2} &=& - (R^i_{\ 0 j 0} - \partial_j w^i ) \eta^j + \frac{1}{\ell^2}\eta^j \partial_j F^i,
\end{eqnarray}
and hence we can reformulate Eq.~(\ref{til33}) by introducing  a newly defined tidal force vector $\mathcal{F}^i$, and the tidal matrix $K^i_j$, which has been modified by the matter-curvature coupling and Weyl geometry,
\begin{eqnarray}
	\frac{d^2 \eta^i}{dt^2} &=& \mathcal{F}^i = K^i_j \eta^j,
\end{eqnarray}
with the explicit expression of $K^i_j$ given by
\begin{eqnarray}
	K^i_j &=& - R^i_{\ 0 j 0} + \partial_j w^i + \frac{1}{\ell^2} \partial_j F^i,
\end{eqnarray}
where $F^i$ is the component of extra force, and has been defined in Eq.(\ref{mm}). The contraction of the tidal matrix gives,
\begin{eqnarray}
	K &=& K^i_i = \frac{\partial F^i}{\partial \eta^i}
	= - R_{00} + \partial_i w^i + \frac{1}{\ell^2} \partial_i F^i,
\end{eqnarray}
and by using Eq.~(\ref{til24}), (\ref{til11}, and (\ref{til10}), we obtain
\begin{eqnarray}\label{til31}
	K^i_j &=& - \frac{\partial^2 \Phi}{\partial x^i x^j} + \partial_j w^i  + \frac{1}{\ell^2} \partial_j F^i,
\end{eqnarray}
and the scalar $K$ reads,
\begin{eqnarray}
	K &=& - \Delta \Phi + \partial_i w^i + \frac{1}{\ell^2} \partial_i F^i,
\end{eqnarray}
Substituting the expression \eqref{mm} for $F^i$, one obtains
\begin{align}
K^i_j &= - \frac{\partial^2 \Phi}{\partial x^i x^j} -2\kappa^2\partial_j\left(\frac{\partial_i f_T}{1+2\kappa^2 f_T}\right),\nn
\end{align}
and
\bea
K &=& - \Delta \Phi -\frac{2\kappa^2}{(1+2\kappa^2 f_T)^2}\times \nonumber \\
&&\Big[(1+2\kappa^2 f_T)\Box f_T-2\kappa^2\partial_i f_T\partial_i f_T\Big],
\eea
respectively.

\subsection{The Roche radius in Weyl type $f(Q,T)$ gravity}

In Newtonian gravity, the spherical potential of a given particle with mass $M$ is given by
\begin{eqnarray}\label{roch1}
\Phi(r) = -\frac{M}{8\pi r},
\end{eqnarray}
where we have assumed that $8\pi G=1$. In a frame of reference with the $x$ axis passing through the particle's position, indicating that the particle is located at  ($ x=r, y=0, z=0$), the Newtonian tidal force tensor $\tau_{ij}$ will be diagonal, and has only the following nonzero components \cite{harko2012},
\begin{eqnarray}\label{roch2}
\tau_{ij} = - \partial_i \partial_j \Phi = diag\bigg(\frac{2M}{8\pi r^3}, -\frac{M}{8\pi r^3}, -\frac{M}{8\pi r^3}\bigg).
\end{eqnarray}
The components of the Newtonian tidal force $\mathcal F_i$ can be written as
\begin{eqnarray}\label{roch3}
\hspace{-0.5cm}\mathcal F_x = \frac{2M\Delta m\, x}{8\pi r^3}, \mathcal F_y = - \frac{M\Delta m \,y}{8\pi r^3}, \mathcal F_z = - \frac{M \Delta m\, z}{8\pi r^3},
\end{eqnarray}
respectively \cite{hans}.

The Roche limit, an important astrophysical and astronomical concept, is defined as the closest distance $r_{Roche}$ that a cosmic object, having mass $m$, radius $R_m$, and density $\rho_m$, respectively, can come near a massive star of mass $M$, radius $R_M$ and density $\rho_M$, respectively, without being torn apart by the tidal gravity of the star. In the following we consider a simplified case with $M \gg m$, a condition which allows us to set the center of mass in the geometrical center of the mass  $M$.

We consider a small object of mass $\Delta m$ located at the surface of the small body of mass $m$. The gravitation force from the small mass acting on $\Delta m$ is given by
\begin{eqnarray}\label{roch4}
F_G = \frac{m \Delta m}{8\pi R_m^2},
\end{eqnarray}
while the tidal force from the big massive body acting on $\Delta m$ is obtained as
\begin{eqnarray}\label{roch5}
\mathcal F = \frac{2M\Delta m R_m}{8\pi r^3},
\end{eqnarray}
where $r$ is the distance between the centers of the two celestial objects, and we have neglected the differences in the distances between $\Delta m$ and $M$,  and $\Delta M$ and $m$, respectively. The Roche limit is reached if the two forces acting on $\Delta m$ are equal, $F_G = \mathcal F$. Thus we obtain the Roche limit in Newtonian gravity $r_{Roche}$ as
\begin{eqnarray}\label{roch6}
r_{Roche} = R_m \bigg(\frac{2M}{m}\bigg)^{\frac{1}{3}} = 2^{\frac{1}{3}} R_M \bigg(\frac{\rho_M}{\rho_m}\bigg)^{\frac{1}{3}}.
\end{eqnarray}

With the use of Eq.~(\ref{nl11}) we obtain the modification of the gravitational force in Weyl-type $f(Q,T)$ gravity, which can be represented as
\begin{align}\label{roch8}
F_{total} &= F_{gravity} + F_{geometry} +F_{ Extraforce}\nn
&= F_{gravity} -\f{M\ell}{8\pi R_m^2}+ \ell w^r + \frac{F^r}{\ell}+\frac{(\ell^2)'}{2\ell^2} u^r,
\end{align}
where the index $r$ indicates the radial components, $F_{gravity}$ is the Newtonian gravitational force,  $F_{geometry}$ gives by modifications from geometry components beyond Riemann geometry, and $F_{ Extraforce}$ is the component generated by the geometry-matter coupling.

Thus, in Weyl geometry, by using Eq.~(\ref{til31}), the Roche limit $r_{Roche}$ is obtained as
\begin{align}\label{roch11}
& \Bigg[ \frac{2M}{8\pi r^3_{Roche}} + \partial_r w^r + 2w_r^{2} + \frac{1}{\ell^2} \partial_r F^r \Bigg] R_m\nn
& = \frac{m}{8\pi R_m^2} -\f{M\ell}{8\pi R_m^2}+ \ell w^r + \frac{F^r}{\ell}+\frac{(\ell^2)'}{2\ell^2} u^r,
\end{align}
and the vectors containing index $r$ (no summation upon $r$) must be evaluated in the coordinate system in which the Newtonian tidal tensor is diagonal. Considering that the gravitational effects due to the coupling between matter and curvature are small as compared to the Newtonian ones, we have
\begin{eqnarray}\label{roch12}
r_{Roche} &\approx& R_m \bigg(\frac{2M}{m}\bigg)^{\frac{1}{3}}\times\nn
&& \Bigg[1 +\f{M\ell}{3m}+ \frac{8\pi R_m^3}{3 m} \bigg(\partial_r w^r + \frac{1}{\ell^2} \partial_r F^r\bigg) -\nn
&& \frac{8\pi R_m^2}{3 m}\bigg(\ell w^r + \frac{F^r}{\ell}+\frac{(\ell^2)'}{2\ell^2} u^r\bigg)\Bigg].\nn
\end{eqnarray}

Substituting $F^r$ from Eq.~\eqref{mm} one finally obtains
\begin{align}
r_{Roche} &\approx R_m \bigg(\frac{2M}{m}\bigg)^{\frac{1}{3}}\times\nn
& \Bigg[1 +\f{M\ell}{3m}+ \frac{16\pi \kappa^2R_m^3}{3 m} \partial_r\left(\f{h^{rr}\partial_r f_T}{1+2\kappa^2 f_T}\right) +\nn
&\frac{8\pi R_m^2}{3 m}\bigg(\f{2\kappa^2\ell}{1+2\kappa^2f_T}h^{rr}\partial_rf_T-\frac{(\ell^2)'}{2\ell^2} u^r\bigg)\Bigg].
\end{align}

\section{Discussions and final remarks}\label{sect5}

Abandoning the metricity conditions, and including nonminimal curvature-matter couplings are some promising ways to modify standard general relativity, and to explain the major challenges present day gravity theories face. A possible geometric avenue for the generalization of general relativity is represented by the so-called symmetric teleparallel gravity theory \cite{nester1999}, and by its extensions \cite{beltran2018,fQ1,fQ2}. In particular, the role of matter and of the geometry-matter couplings have been analyzed in \cite{fQT} and \cite{Xu2020}, respectively.  In the present paper we have extended the previous analyses of the Weyl type $f(Q,T)$ gravity, a particular version of the general $f(Q,T)$ type theories, by developing  some basic theoretical tools that would allow not only to further investigations of the fundamental geometrical and physical properties of these  gravity theories, but can also open the possibility of their observational testing.

More exactly, from the fundamental point of view of the analysis of the Weyl type $f(Q,T)$  theories, we have obtained two of the basic equations of the gravitational physics, namely, the geodesic deviation equation, and the Raychaudhuri equation, respectively. The geodesic deviation equation geodesic describes the way objects approach or recede from one another when moving under the influence of a spatially varying gravitational field. One of the important applications of the geodesic deviation equation is in the study of the tidal forces, which in  modified theories of gravity acquire some extra terms due to the presence of the new terms that modify the gravitational interaction. Hence the geodesic equation can be used to observationally test the Weyl type $f(Q,T)$ gravity model through the observations of the  effects of the tides produced by an extended mass distribution. Tidal effects play an important role in the eccentric inspiralling neutron star binaries \cite{TF1}. The neutron stars can be modelled as a compressible ellipsoid, which can deform nonlinearly due to tidal forces, while the orbit evolution can usually be described with the post-Newtonian  theory. In general, the tidal interaction can accelerate the inspiral, and cause orbital frequency and phase shifts. Tidal interactions have an essential effect on the star formation in galaxies, since tidal perturbations induced by close companions increase the gas accretion rates \cite{TF2}. By using gravitational wave detector networks one can constrain the equation of state of binary neutron-stars, and extract their redshifts through the imprints of tidal effects in the gravitational waveforms \cite{TF3}. The existence of light, fundamental bosonic fields is an attractive possibility that can be tested via black hole observations. The effect of a tidal field  caused by a companion star or black hole on the evolution of superradiant scalar-field states around spinning black holes can test the existence of light bosonic fields \cite{TF4}. For large tidal fields the scalar condensates are disrupted, and the impact of tides can be relevant for known black-hole systems such as the one at the center of our galaxy or the Cygnus X-1 system. The companion of Cygnus X-1 will disrupt possible scalar structures around the black hole for large gravitational couplings. Tidal effects in massless scalar-tensor theories were considered in \cite{TF5}, where  a new class of scalar-type tidal Love numbers. It turns out that in a system dominated by dipolar emission, tidal effects may be detectable by LISA or third generation gravitational wave detectors.  Another astrophysical situations in which the effects
of the tides are of major importance  are perturbations of the Oort cloud by the
galactic field, globular clusters evolving under the influence of the galactic mass distribution, and galactic encounters \cite{TF6}.  As we have seen in our analysis of the $F(q,t)$ gravity, the curvature-matter coupling significantly modifies the nature of the tidal forces, as well as the equation of motion in the Newtonian limit. Therefore, the comparison of the theoretical predictions of the Weyl type $f(Q,T)$ gravity about the modifications of the tidal forces with the observational evidences, coming from a large class of astrophysical phenomena could give, at least in principle, some insights into the fundamental aspects of the gravitational interaction, and its geometric description.

We have also obtained the generalization of the Poisson equation, describing the properties of the gravitational potential. The Poisson equation, and its solution, is an important tool in the investigation of many gravitational effects involving small velocities and low matter densities. The modifications of the gravitational potential, and the new terms appearing in the equation may provide a theoretical explanation for the observed dynamics of the particles moving on circular orbits around galaxies. These observations are usually explained by postulating the existence of dark matter, a mysterious major component of the Universe, which has not been detected yet. Hence the novel geometric effects induced by the Weyl-type $f(Q,T)$ gravity may provide a geometric explanation for the galactic dynamics of test particles without having to resort to the dark matter hypothesis. In present paper we have obtained the equation of motion of the particles in the Weyl-type $f(Q,T)$ gravity, and we have discussed it in detail. Note that the extra force has two components of different origins, coming from the matter distribution and from the geometrical properties of the space-time. The geometry provides an extra degree of freedom, with the nonminimal curvature-matter coupling also generating more degrees of freedom for the gravitational interaction. Hence these extra degrees of freedom contribute with new terms to the extra force, the tidal force, and the Roche limit. These extra terms may have observational (and even experimental) effects, which can be used to test the theoretical gravity model we have investigated in this paper.

 The Raychaudhuri equation is of major importance in the investigation of the space-time singularities, and in construction of cosmological models. For the sake of completeness we briefly mention some cosmological applications of our results. Let's consider a flat Friedmann-Lemaitre-Robertson-Walker Universe, with metric given by $d s^2 = - dt^2 + a^2 (t) \delta_{ij} dx^i dx^j$, where $a(t)$ is the scale factor. We  take $\ell = 1$, and we adopt a co-moving reference system, with $u^\mu = (1,0,0,0)$. In this case the prime operator is given by ${}' = u^\mu \tn_\mu = u^0 \tn_0$, and, when applied on a scalar, we have ${}' = d/d\tau = d/dt$. We also introduce  the Hubble function $H=(1/a(t))da(t)/dt$, describing the rate of change of $a(t)$ with respect to time.

We consider a general model for the cosmological nonmetricity, which was introduced in \cite{iosifidis2020}, and according to which
$Q_{\lambda\mu\nu} = A(t) u_\lambda h_{\mu\nu} + B(t) h_{\lambda(\mu} u_{\nu)} + C(t) u_\lambda u_\mu u_\nu$,
where $A$, $B$, $C$ are time-dependent functions representing the behaviour of non-Riemannian degrees of freedom and $h_{\mu\nu}$ is the projection tensor previously defined. With the Weyl vector defined in Eq.~(\ref{weyl12}), only one extra degrees of freedom is added by the nonmetricity, By assuming  $A = - C$, and by taking $B = 0$, we obtain for the cosmological nonmetricity the expression
\begin{eqnarray}\label{flrw9}
	&& Q_{\lambda\mu\nu} = A u_\lambda g_{\mu\nu} = -2 w_\lambda g_{\mu\nu},
\end{eqnarray}
giving for the Weyl vector the simple expression
\begin{eqnarray}\label{flrw21}
	w_\lambda = - \frac{A}{2} u_\lambda
\end{eqnarray}

By taking advantage of Eq.(\ref{nl17}), we obtain for the extra force the expression,
\begin{eqnarray}
f^\rho &=& F^\rho - \frac{A}{2} u^\rho.
\end{eqnarray}
We also notice that under conditions that we took formerly in this approach, from a geometrical point of view we have,
\begin{eqnarray}\label{flrw15}
f^\rho = u^\mu \tn_\mu u^\rho = -\frac{A}{2} u^\rho,
\end{eqnarray}
leading to $F^\rho = 0$. This condition is a direct consequence of the use of the co-moving reference frame. In the laboratory reference system the extra-force does not vanish.

If we keep the terms containing $F^\rho$, the Raychaudhuri equation in the Weyl spacetime, given by Eq.~(\ref{ray14}), can now be written as (see Appendix \ref{app8} for the calculational details),
\begin{eqnarray}\label{flrw20}
3H' + 3H^2 &=& \frac{3A'}{2} + \frac{3A}{2} H - \tilde R_{\mu\nu}u^\mu u^\nu +\tn_\mu F^\mu,\nn
\end{eqnarray}
and it can also be reformulated into a equation of $a(t)$,
\begin{eqnarray}
\frac{a''}{a} =\frac{A'}{2} + \frac{Aa'}{2a} -\f13 \tilde R_{\mu\nu}u^\mu u^\nu + \tn_\mu F^\mu,
\end{eqnarray}
which represents the generalized cosmic acceleration equation. Considering the fact that $F^\rho=0$ in the adopted coordinate system,  and after substituting the expression of $\tilde R_{\mu\nu}u^\mu u^\nu$, one can see that the above equation is identically satisfied. The most general form of the acceleration equation in the presence of torsion and non-metricity has been obtained, in a Friedmann-lemaitre-Robertson-Walker geometry,  with the help of the generalized Raychaudhuri equation with torsion and non-metricity, in \cite{iosifidis2020}. Cosmological Hyperfluids, representing fluids with intrinsic hypermomentum that induce spacetime torsion and non-metricity, were studied in \cite{iosifidis2020a}, where the most general form of the Friedmann equations with torsion and non-metricity were also obtained.

To conclude, the present investigation of some fundamental  aspects of the Weyl type $f(Q,T)$ gravity opens further possibilities for the theoretical, observational and even experimental study of the alternative purely geometrical theories of gravity in the presence of the coupling between geometry and matter. Moreover, the results obtained in the present paper may also be relevant for other classes of modified gravity theories.

\section*{Acknowledgments}

We would like to thank the two anonymous referees for comments and suggestions that helped us to improve our manuscript. T. H. thanks the Yat Sen School of the Sun Yat Sen University in Guangzhou, P. R. China,  for the warm hospitality offered during the preparation of this work.

\appendix

\section{Calculational details for some basic results}\label{app}

In this Appendix we present explicitly some of the calculational details and intermediate steps that were used in the derivation of the basic results in the main text of our paper.

\subsection{Obtaining the  expression of the scalar nonmetricity $Q$}\label{app1}

One can obtain $L^\lambda_{\mu\nu}$ explicitly in Weyl geometry by its definition  Eq.(\ref{weyl14}), as
\begin{eqnarray}\label{ap1}
L^\nu_{\mu\nu} = w^\lambda g_{\mu\nu} - w_\mu \delta^\lambda_\nu - w_\nu \delta^\lambda_\mu.
\end{eqnarray}
Then with the use of Eq.(\ref{weyl13}) we obtain the scalar non-metricity $Q$ as follows,
\begin{eqnarray}
Q &=& -g^{\mu\nu}\bigg[\l(w_\beta \delta^\alpha_\mu + w_\mu \delta^\alpha_\beta - w^\alpha g_{\beta\mu}\r)\times\nn
&&\l(w_\alpha \delta^\beta_\nu + w_\nu \delta^\beta_\alpha - w^\beta g_{\nu\alpha}\r)-\nn
&&\l(w_\beta\delta^\alpha_\alpha + w_\alpha \delta^\alpha_\beta - w^\alpha g_{\alpha\beta}\r)\times\nn
&&\l(w_\mu \delta^\beta_\nu + w_\nu \delta^\beta_\mu - w^\beta g_{\mu\nu}\r)\bigg]
= 6w_\mu w^\mu.
\end{eqnarray}

\subsection{The explicit expressions of the covariant derivative in Weyl geometry}\label{app2}

We will present below the explicit expression of the covariant derivative in Weyl geometry of certain types of tensors. For vectors,
\begin{eqnarray}
\tn^\mu A^\lambda &=& g^{\mu\nu} \tn_\nu A^\lambda = g^{\mu\nu} (\partial_\nu A^\lambda + \tg^\lambda_{\sigma\nu}A^{\sigma}) \nn
&=& \nabla^\mu A^\lambda + w^\lambda A^\mu - w^\mu A^\lambda - g^{\mu\lambda} w_\sigma A^\sigma,\\
\tn^\mu A_\lambda &=& g^{\mu\nu} \tn_\nu A_\lambda = g^{\mu\nu} (\partial_\nu A_\lambda - \tg^\sigma_{\lambda\nu}A_{\sigma}) \nn
&=& \nabla^\mu A_\lambda - \delta^\mu_\lambda w^\sigma A_\sigma + w^\mu A_\lambda + w_\lambda A^\mu.
\end{eqnarray}
For second order tensors we obtain,
\begin{eqnarray}
\tn^\lambda A^{\mu\nu} &=& g^{\lambda\sigma} \tn_\sigma A^{\mu\nu}\nonumber\\
&=& g^{\lambda\sigma} (\partial_\sigma A^{\mu\nu} + \tg^\mu_{\sigma\alpha} A^{\alpha\nu} + \tg^\nu_{\sigma\alpha}A^{\mu\alpha})\nn
&=& \nabla^\lambda A^{\mu\nu} + w^\mu A^{\lambda\nu} + w^\nu A^{\mu\lambda} -\nn
&& g^{\lambda\mu} w_\alpha A^{\alpha\nu} - g^{\lambda\nu} w_\alpha A^{\mu\alpha} - 2w^\lambda A^{\mu\nu},\\
\tn^\lambda A_{\mu\nu} &=& g^{\lambda\sigma} \tn_\sigma A_{\mu\nu}\nn
&=& g^{\lambda\sigma} (\partial_\sigma A_{\mu\nu} - \tg^\alpha_{\sigma\mu} A_{\alpha\nu} - \tg^\alpha_{\sigma\nu}A_{\mu\alpha})\nn
&=& \nabla^\lambda A_{\mu\nu} - \delta^\lambda_\mu w^\alpha A_{\alpha\nu} - \delta^\lambda_\nu w^\alpha A_{\mu\alpha} +\nn
&& w_\mu A^\lambda_{\ \nu} + w_\nu A_\mu^{\ \lambda} + 2 w^\lambda A_{\mu\nu},\\\nn
\tn^\lambda A^\mu_{\ \nu} &=& \tn^\lambda (g_{\sigma\nu} A^{\mu\sigma})
= Q^\lambda_{\ \sigma\nu} A^{\mu\sigma} + g_{\sigma\nu} \tn^\lambda A^{\mu\sigma}\nn
&=& \nabla^\lambda A^\mu_{\ \nu} + w^\mu A^\lambda_{\ \nu} + w_\nu A^{\mu\lambda} -\nn
&& g^{\lambda\mu} w_\alpha A^\alpha_{\ \nu} - \delta^\lambda_\nu w_\alpha A^{\mu\alpha} - 2w^\lambda A^\mu_{\ \nu}.
\end{eqnarray}

\subsection{The perfect fluid model in Weyl geometry}\label{app3}

We should first notice that since the magnitude of the four-velocity in Weyl geometry is not preserved, and $u_\mu u^\mu = -\ell^2$, the perfect fluid model must be generalized to include the effects $\ell$. Generally, a perfect fluid can be characterized by its four-velocity $u$, and the thermodynamic  quantities - the proper density $\rho$, the isotropic pressure $p$, the temperature $T$, and the specific entropy $s$, or the specific enthalpy $\omega = (p+\rho)/n$ \cite{gron} (in this part $\omega$ is defined independently from the main body). Here $n$ is the conserved baryon number density, and $n$ does not change its magnitude during parallel transport. We also introduce the particle number density four-vector $n^{\mu}$, defined as
\begin{equation}\label{ttt1}
n^\mu = n\sqrt{-g} u^\mu.
\end{equation}
Consequently,
\begin{equation}\label{ttt2}
n = \sqrt{\frac{g_{\mu\nu} n^\mu n^\nu}{\ell^2 g}},
\end{equation}
where $g$ is the determinant of the matrix $g_{\mu\nu}$. Next we need to introduce the matter Lagrangian, which we assume as depending only on the energy density scalar the in local rest frame of the fluid,
\begin{equation}
\cL_m = -\rho.
\end{equation}

The energy-momentum tensor of the fluid is given by $T_{\mu\nu} = -2\partial \cL_m/\partial g^{\mu\nu} + g_{\mu\nu} \cL_m$ under the constraints \cite{gron}
\begin{eqnarray}
\delta s = 0,\quad \delta n^\mu = 0.
\end{eqnarray}

From the thermodynamical relation,
\begin{equation}\label{ttt3}
\l(\frac{\partial \rho}{\partial n}\r) = \omega,
\end{equation}
we immediately obtain $\delta \rho = \omega \delta n$.

Using Eqs.~(\ref{ttt1},\ref{ttt2},\ref{ttt3}), we obtain
\begin{eqnarray}
\delta n &=& \frac{1}{2n} \bigg(\frac{n^\mu n^\nu}{\ell^2 g} \delta g_{\mu\nu} - n^\mu n^\nu \frac{g_{\mu\nu}}{\ell^2 g^2} \delta g \bigg)\nn
&=& \frac{n}{2}\bigg(-\frac{u^\mu u^\nu}{\ell^2} \delta g_{\mu\nu} + \frac{u^\mu u_\mu}{\ell^2 g}\delta g\bigg).
\end{eqnarray}
By using the basic properties of the metric variation,
\begin{eqnarray}
&& \delta g_{\mu\nu} = - g_{\mu\lambda} g_{\nu\sigma} \delta g^{\lambda\sigma},\\
&& \frac{\delta g}{\delta g^{\mu\nu}} = -g_{\alpha\mu}g_{\beta\nu} \frac{\delta g}{\delta g_{\alpha\beta}} = -gg_{\mu\nu},
\end{eqnarray}
we obtain
\begin{equation}
\delta n = \frac{n}{2}\bigg(\frac{u_\mu u_\nu}{\ell^2} + g_{\mu\nu}\bigg)\delta g^{\mu\nu}.
\end{equation}
The derivative of the matter Lagrangian with respect to the metric tensor is given by
\begin{equation}
\frac{\partial \cL_m}{\partial g^{\mu\nu}} = -\frac{n\omega}{2}\bigg(\frac{u_\mu u_\nu}{\ell^2} + g_{\mu\nu}\bigg),
\end{equation}
and hence we finally obtain the energy-momentum tensor of the perfect fluid model in Weyl geometry as,
\begin{equation}
T_{\mu\nu} = \frac{p+\rho}{\ell^2} u_\mu u_\nu + p g_{\mu\nu}.
\end{equation}

\subsection{The details of the calculation of the extra force in Weyl geometry}\label{app4}

The divergence of the matter energy-momentum tensor in Weyl geometry is obtained as follows. We first obtain
\begin{eqnarray}
&& \tn^\mu T_{\mu\nu}= \nabla^\mu T_{\mu\nu} - 2w^\mu T_{\mu\nu} + w_\mu T\nn
&=& \frac{\tn^\mu p + \tn^\mu \rho}{\ell^2} u_\mu u_\nu - \frac{p+\rho}{\ell^4} u_\mu u_\nu \tn^\mu \ell^2 +\nn
&& \frac{p+\rho}{\ell^2} (u_\mu \tn^\mu u_\nu + u_\nu \tn^\mu u_\mu) + g_{\mu\nu} \tn^\mu p + p \tn^\mu g_{\mu\nu}\nn
&=& \frac{p' + \rho'}{\ell^2} u_\nu - \frac{p+\rho}{\ell^4} (\ell^2)' u_\nu +\nn
&& \frac{p+\rho}{\ell^2} (a_\nu + u_\nu \tn^\mu u_\mu) + \tn_\nu p + p Q^\mu_{\ \mu\nu}.
\end{eqnarray}
Then after multiplication with the projection tensor we have
\begin{eqnarray}
&& h^{\rho\nu} \tn^\mu T_{\mu\nu}
= h^{\nu\rho} \bigg(\frac{p+\rho}{\ell^2} a_\nu + \tn_\nu p + pQ^\mu_{\ \mu\nu}\bigg)\nn
&=& \frac{p+\rho}{\ell^2} (A^\rho - Q^{\mu\nu\rho} u_\mu u_\nu) + \frac{p+\rho}{\ell^4} a_\mu u^\mu u^\rho +\nn
&& h^{\rho\nu} \tn_\nu p + h^{\nu\rho} Q^\mu_{\ \mu\nu} p\nn
&=& \frac{p+\rho}{\ell^2} (A^\rho - Q^{\mu\nu\rho} u_\mu u_\nu) -\nn
&& \frac{p+\rho}{2\ell^4} \bigg[(\ell^2)' + Q^{\mu\nu\lambda} u_\mu u_\nu u_\lambda\bigg] u^\rho + h^{\rho\nu} \tn_\nu p + h^{\nu\rho} Q^\mu_{\ \mu\nu} p.\nn
\end{eqnarray}

\subsection{Calculational details of the  derivation of the geodesic deviation equation}\label{app5}

 The total derivative in Weyl geometry of the term $u^\nu\tilde\nabla_\nu U^\mu $ is given by,
\begin{eqnarray}
u^\nu\tilde\nabla_\nu U^\mu = u^\nu(\nabla_\nu U^\mu + w^\mu U_\nu - w_\nu U^\mu - \delta^\mu_\nu w_\alpha U^\alpha),\nn
\end{eqnarray}
and
\begin{eqnarray}
\hspace{-0.8cm}U^\nu\tilde\nabla_\nu u^\mu = U^\nu(\nabla_\nu u^\mu + w^\mu u_\nu - w_\nu u^\mu - \delta^\mu_\nu w_\alpha u^\alpha),
\end{eqnarray}
respectively. Note that from the definition $\partial U^\mu /\partial \lambda = \partial u^\mu /\partial \sigma$ and $w_\nu U^{[\mu}u^{\nu]} = 0$, in Weyl space we obtain $u^\nu\tilde\nabla_\nu U^\mu = U^\nu\tilde\nabla_\nu u^\mu$.

\subsection{Details of the calculations in the derivation of the Raychaudhuri equation}

Some important steps an intermediate results in the derivation of the Raychaudhury equation are as follows:
\begin{eqnarray}
\theta &=& g^{\mu\nu} \tn_\nu u_\mu = \nabla^\mu u_\mu - 2w_\mu u^\mu,
\end{eqnarray}
\begin{eqnarray}
a_\mu &=& u^\alpha \nabla_\alpha u_\mu - \ell^2 w_\mu = f_\mu + 2w_\nu u^\nu u_\mu,
\end{eqnarray}
\begin{eqnarray}
\xi_\mu &=& u^\alpha\nabla_\mu u_\alpha- \ell^2 w_\mu,
\end{eqnarray}
\begin{eqnarray}
\hspace{-0.5cm}\nabla_\mu (u_\alpha u^\alpha) = - \nabla_\mu (\ell^2) = 2u_\alpha \nabla_\mu u^\alpha = 2u^\alpha \nabla_\mu u_\alpha,
\end{eqnarray}
\begin{eqnarray}
\sigma_{\mu\nu} &=& \frac{1}{2}(\nabla_\nu u_\mu + \nabla_\mu u_\nu) - \frac{1}{3} \nabla_\alpha u^\alpha g_{\mu\nu} +\nn
&& \frac{2}{3\ell^4} u_\mu u_\nu u^\alpha u^\beta \nabla_\alpha u_\beta + \frac{1}{2\ell^2} (u_\nu u^\alpha \nabla_\alpha u_\mu +\nn
&& u_\mu u^\alpha \nabla_\alpha u_\nu + u_\mu u^\alpha \nabla_\nu u_\alpha + u_\nu u^\alpha \nabla_\mu u_\alpha) -\nn
&& \frac{1}{3\ell^2} (u^\alpha u^\beta \nabla_\alpha u_\beta g_{\mu\nu} + u_\mu u_\nu \nabla_\alpha u^\alpha),
\end{eqnarray}
\begin{eqnarray}
\hspace{-0.3cm}\omega_{\mu\nu} &=& \frac{1}{2}(\nabla_\nu u_\mu - \nabla_\mu u_\nu) + \frac{1}{2\ell^2}(u_\nu u^\alpha \nabla_\alpha u_\mu -\nn
&& u_\mu u^\alpha \nabla_\alpha u_\nu + u_\mu u^\alpha \nabla_\nu u_\alpha - u_\nu u^\alpha \nabla_\mu u_\alpha).
\end{eqnarray}
\\
\subsection{The Riemann and Ricci tensors in the weak-field approximation}\label{app7}

In the weak field approximation the Riemann and the Ricci tensors can be obtained as follows
\bea\label{nl30}
 R^\alpha_{\ 0 \beta 0} = g^{\alpha\rho} R_{\rho 0 \beta 0} = R^i_{~ 0 l 0},
\eea
\bea\label{nl31}
R^0_{\ \alpha 0 \beta} = g^{0 \rho} R_{\rho \alpha 0 \beta} =-R_{\alpha 0 \beta 0},
\eea
\bea\label{nl32}
R_{lm} =  R^0_{\ l 0 m} + \delta^j_i R^i_{\ l j m} = - R_{l 0 m 0} + R^i_{\ l i m}.
\eea
\\
\subsection{Calculations of the cosmological terms in the flat FLRW spacetime}\label{app8}

In a flat FLRW geometry the expressions of the relevant cosmological quantities can be obtained as follows:
\begin{eqnarray}
\nabla_\mu w^\mu  = \frac{1}{\sqrt{-g}} \frac{\partial(\sqrt{-g} w^0)}{\partial x^0} = -\frac{A'}{2} - \frac{3A}{2} H,
\end{eqnarray}
\begin{eqnarray}
\theta &=& \frac{1}{\sqrt{-g}} \frac{\partial(\sqrt{-g} u^0)}{\partial x^0} -A = 3 H - A,
\end{eqnarray}
\begin{eqnarray}
R_{\mu\nu} u^\mu u^\nu = R_{00} (u^0)^2 = -3 (H^\prime + H^2),
\end{eqnarray}
\begin{eqnarray}
W_{\mu\nu} &=& \nabla_\nu w_\mu - \nabla_\mu w_\nu = \partial_\nu w_\mu - \partial_\mu w_\nu = 0,
\end{eqnarray}
\begin{eqnarray}
\sigma_{\mu\nu} = 0 = \omega_{\mu\nu} ,
\end{eqnarray}
\begin{eqnarray}
\Gamma^\mu _{\ \mu 0} &=& \frac{1}{2} g^{\mu\nu} (g_{\nu\mu ,0}+g_{\nu 0 ,\mu} - g_{\mu 0 ,\nu})\nn
&=& \frac{1}{2} g^{\mu\mu} (g_{\mu\mu ,0}+g_{\mu 0 ,\mu} - g_{\mu 0 ,\mu})\nonumber\\
&=& \frac{1}{2} g^{\mu\mu} \partial_0 g_{\mu\mu}.
\end{eqnarray}
\end{document}